\documentclass[twocolumn]{aastex631}  %twocolumn
\usepackage{enumitem}
\usepackage{xcolor}
\usepackage{xcolor, soul}
\sethlcolor{yellow}
\usepackage{subfigure}
\usepackage{tablefootnote}
\usepackage{bm}
\usepackage{lineno}
\hypersetup{backref=true,pagebackref=true,  hyperindex=true,colorlinks=blue,breaklinks=true,  urlcolor=violet,linkcolor= red, bookmarks=true, 	bookmarksopen=true,  filecolor=cyan, citecolor=blue, linkbordercolor=blue}
\shorttitle{Self-Lensing/Occultation Signals due to WDMS binary systems}
\shortauthors{Sajadian, Fatheddin}

\begin{document}
%\linenumbers
\title{Finite-Lens Effect on Self-Lensing in detached White Dwarfs-Main Sequence Binary Systems}

\author[0000-0002-0167-3595]{Sedighe Sajadian}
\affiliation{Department of Physics, Isfahan University of Technology, Isfahan 84156-83111, Iran, \url{s.sajadian@iut.ac.ir}}

\author[0000-0002-7611-9249]{Hossein Fatheddin} 
\affiliation{Leiden Observatory, Leiden University, PO Box 9513, NL-2300 RA Leiden, the Netherlands}

\begin{abstract}
In an edge-on and detached binary system, including a white dwarf (WD) and a main-sequence star (WDMS), when the source star is passing behind the compact companion its light is bent and magnified. Meanwhile, some part of its images' area is obscured by the WD's disk. These two effects occur simultaneously, and the observer receives the stellar light magnified and partially obscured due to the finite-lens size. We study these effects in different WDMS binary systems numerically using inverse-ray-shooting (IRS) and analytically using approximate relations close to reality. For WDMS systems with long orbital periods $\gtrsim 300$ days and $M_{\rm{WD}}\gtrsim 0.2 M_{\sun}$ ($M_{\rm{WD}}$ is the mass of WD), lensing effects dominate the occultations due to finite-lens effects, and for massive WDs with masses higher than solar mass no occultation happens. The occultations dominate self-lensing signals in systems with low-mass WDs($M_{\rm WD}\lesssim 0.2 M_{\sun}$) in close orbits with short orbital periods $T\lesssim 50$ days. The occultation and self-lensing cancel each other out when the WD's radius equals $\sqrt{2}$ times the Einstein radius, regardless of the source radius, which offers a decreasing relation between the orbital periods and WDs' mass. We evaluate the errors in maximum deviations in self-lensing/occultation normalized flux which are made by using its known analytical relation and conclude that these errors could be up to $0.002,~0.08,~0.03$ when the orbital period is $T=30,~100,~300$ days, respectively. The size of stellar companions in WDMSs has a twofold manner as it decreases the depth of self-lensing/occultation signals but enlarges their width. 
\end{abstract}

\keywords{Astronomical simulations -- Compact binary stars -- Stellar remnants -- Compact objects -- Gravitational lensing}

%%%%%%%%%%%%%%%%%%%%%%%%%%%%%%%%%
\section{Introduction}\label{sec2}
Gravitational lensing refers to the bending of light due to passing from the gravitational field of a massive object\citep[see, e.g., ][]{soldner1921distraction}. The correct amount of the deflection angle in a lensing event was derived from the General Theory of Relativity \citep[e.g., ][]{clark1972uniform,renn1997origin}. There are several kinds of gravitational lensing depending on the lens mass, the impact parameter, and the distances between the lens, source object, and the observer \citep{1992bookSchneider,1996Narayan}. The first one is gravitational microlensing which refers to a temporary enhancement in the brightness of a background star which is collinear with a foreground and massive object \citep{Einstein1936, Liebes1964, 1964MNRASrefsdal}. This kind of lensing is a method for discovering and characterizing isolated and even dark objects ranging from extrasolar planets to isolated stellar-mass black holes in our galaxy \citep[e.g., see, ][]{2012Gaudireview,2022ApJSahu,2021MNRAsajadian}. In cosmological scales, either weak or strong lensing happens by generating giant arcs or multiple images from a distant and bright source galaxy. One of the most important applications of these kinds of lensing in cosmological scales is measuring the Hubble constant and studying its variations with the cosmological redshift, and the Hubble tension \citep[e.g., see, ][]{1998AJRiess,2024ApJRiess}.

Gravitational self-lensing is another type of lensing which occurs in detached and edge-on binary systems containing main-sequence stars and compact objects \citep[e.g., ][]{1995ApJGould}. In such systems, whenever the companion star is passing behind the compact object as seen by the observer, its light is bent and magnified due to passing from the gravitational field of the compact object. Self-lensing is a method to detect and characterize compact objects in \textit{detached} binary systems with main-sequence stars. Up to now, five self-lensing events have been discovered from the Kepler data and all of them were related to binary systems including white dwarfs (WDs) and main-sequence stars (WDMS) \citep{KruseAgol2014,2018AJKawahara,2019ApJLMasuda}. Recently, \citet{2024Yamaguchi} revisited the last self-lensing target (KIC 8145411) by high-resolution imaging and found that it was a triplet system including two solar-type stars and one WD with the mass of $\sim 0.53 M_{\odot}$. Therefore, all five discovered self-lensing events were due to WDs more massive than $0.5 M_{\odot}$ in wide orbits with orbital periods of $T\in [88,~728]$ days. Additionally, another self-lensing event (which was first discovered from the Kepler data) was recovered by the Transiting Exoplanet Survey Satellite (TESS\footnote{\url{https://science.nasa.gov/mission/tess/}}, \citet{Ricker2024}) telescope and reported in \citet{2024ApJSorebella}. In the regard of possibility of detecting compact objects through self-lensing/eclipsing signals in the TESS data, \citet{Sajadian2024tess} have done comprehensive simulations and predicted that $15$-$18$ WDs and $6$-$7$ neutron stars (NSs) would be discovered from the TESS Candidate Target List.

To correctly extract the physical parameters of compact objects in self-lensing events, studying their light curves is a crucial step. In self-lensing events, the lens objects and source stars are close to each other and accordingly, their Einstein radii are small in comparison to those in common microlensing events toward the Galactic bulge. The Einstein radius is the radius of the image's ring when the observer, source star and lens object are completely aligned. For instance, in a WDMS binary system whose orbital radius is one astronomical unit (AU), the Einstein radius is $\simeq 0.02$-$0.05R_{\sun}$ whereas for common microlensing events toward the Galactic bulge, it is $R_{\rm E}\sim 1$-$2\rm{AU}$. The lensing events with small Einstein radii are affected by (i) finite-source effect \citep{1994ApJWitt}, because their source radii (projected on the lens plane and normalized to the Einstein radius) are considerable $\rho_{\star}=R_{\star, \rm p}/R_{\rm E} \gtrsim 1$, and (ii) finite-lens effect \citep[see, e.g., ][]{2016ApJHan}. Here, $R_{\star, \rm p}=R_{\star}~D_{\rm l}/\big(D_{\rm l}-x_{\rm o}\big)$ is the source radius ($R_{\star}$) which is projected on the lens plane. $D_{\rm l}$ is the compact object's distance from the observer, and $x_{\rm o}$ is the radial distance of the companion star from the compact object in the line of sight. The second one happens when the size of the lens object is comparable with the Einstein radius that could cover some part of the images' disk. Among different compact objects, WDs have the largest radii with the order of magnitude similar to the Einstein radii in self-lensing events. 

Introducing, studying, and evaluating self-lensing signals in binary systems were done in several papers \citep{1973AAMaeder,1995ApJGould,2002ApJAgol,2016ApJHan}. In this work, we extend these researches and numerically/analytically study some other properties of self-lensing signals affected by finite-lens effects in WDMS binary systems. Here, we will (i) study in what systems occultations due to finite-lens effects are either dominant to or ignorable in comparison with or equal to the self-lensing signals, (ii) evaluate maximum deviations due to occultation, and self-lensing/occultation effects in stellar light curves from different WDMS binary systems, (iii) estimate the errors due to modeling self-lensing peaks using the known analytical relations, and (iv) discuss on the width of self-lensing/occultation signals. 

The paper is organized as follows. In Section \ref{sec3}, we first introduce our formalism for generating self-lensing signals and finite-lens effects in detached and edge-on WDMS binary systems. Then, in Subsection \ref{sec3_2}, we generate the stellar images formed during a self-lensing event and evaluate what WDMS systems have dominant/ignorable occultations (in comparison with self-lensing signals). In Section \ref{sec4}, we simulate and study the properties of self-lensing/occultation stellar light curves. In Subsections \ref{sec4_1}, \ref{sec4_2}, and \ref{sec4_3}, we evaluate (i) the maximum deviations in these stellar light cures based on analytical relations, (ii) the accuracy of the known analytical relations to derive peaks of self-lensing/occultation signals based on the IRS method, and (iii) the widths of self-lensing/occultation signals, respectively. In Section \ref{sec5}, we review the key points and conclusions.  

\section{Self-Lensing and Finite-lens effects}\label{sec3}
\citet{1973AAMaeder} first introduced the lensing effect in different binary systems and concluded the lensing amplitudes in binary systems containing two compact objects would be larger by $10$-$25$ times than those in binary systems including at least one main-sequence star. The probabilities (lensing optical depths) of occurring self-lensing signals in binary systems containing one or two compact objects were estimated by \citet{1995ApJGould}. The analytical and approximate relations for magnification factors during self-lensing events offered by \citet{1973AAMaeder} were extended to non-uniform source stars with a limb-darkening brightness profile by \citet{2002ApJAgol}. Recently, \citet{2016ApJHan} reported a continuous degeneracy while modeling self-lensing/occultation signals that prevents us from uniquely inferring the mass of compact objects. We extend these researches, and study other points.   

In this section, we first introduce our formalism to calculate self-lensing and occultation effects. We then compare the occultation and magnification effects in different WDMS binary systems and determine in what systems the occultation effects dominate self-lensing signals and vise versa.

\subsection{Formalism}\label{sec3_1}
Our formalism to simulate a detached binary system containing a main-sequence star and a WD that are rotating around their common center of mass was described in \citet{Sajadian2024tess} by details and here we review it briefly. To simulate a WDMS binary system, we consider the following parameters: $M_{\star}$ the mass of the main-sequence star, $R_{\star}$ the radius of the main-sequence star, $M_{\rm{WD}}$ the mass of the WD, and $R_{\rm{WD}}$ the radius of the compact object. We also assume the period and eccentricity of their orbits are $T$, and $\epsilon$, respectively. 

If there is no external force and the binary system is isolated, (i) their center of mass is either moving with a constant speed or fixed, and (ii) one component is moving over an ellipse with the same orbital period $T$ concerning the second component. The semi-major axis $a$ of this ellipse can be determined by the Kepler's Third law. 

There are two coordinate systems: (a) the orbital coordinate system, $(x,~y,~z)$, where $z$ is normal to the orbital plane, and $(x,~y)$ describes the orbital plane and are in the directions of major and minor axes of orbital plane, respectively, and (b) the observer coordinate system $(x_{\rm o},~y_{\rm o},~z_{\rm o})$, where $x_{\rm o}$ is toward the observer, $(y_{\rm o},~z_{\rm o})$ are on the sky plane and towards right and up directions, respectively. To convert the first coordinate system to the second one, we need two projection angles: (i) $\theta$ the angle between the semi-minor axis and the sky plane (it is not necessary for circular orbits), and (ii) $i$ the angle between the orbital plane and the line of sight toward the observer, which is the so-called inclination angle.
Accordingly, three components of the position vector of the star with respect to its compact companion in the observer coordinate system are:  
\begin{eqnarray}
x_{\rm o}&=&\cos i~\big(-y \sin \theta + x \cos \theta \big), \nonumber\\
y_{\rm o}&=&y \cos \theta + x \sin \theta, \nonumber\\
z_{\rm o}&=&-\sin i~\big(-y \sin \theta +x \cos \theta \big), 
\end{eqnarray}     
where, $x=a(\cos \xi-\epsilon),$ and $y=a \sin \xi \sqrt{1-\epsilon^{2}}$ describe the stellar position over its orbit in any given time. Here, $\xi$ is the eccentric anomaly and is determined using the Kepler's Equation \citep[see, e.g., ][]{1998Dominik,2017ApJsajadian}. 

\begin{figure*}
\centering
\includegraphics[width=0.32\textwidth]{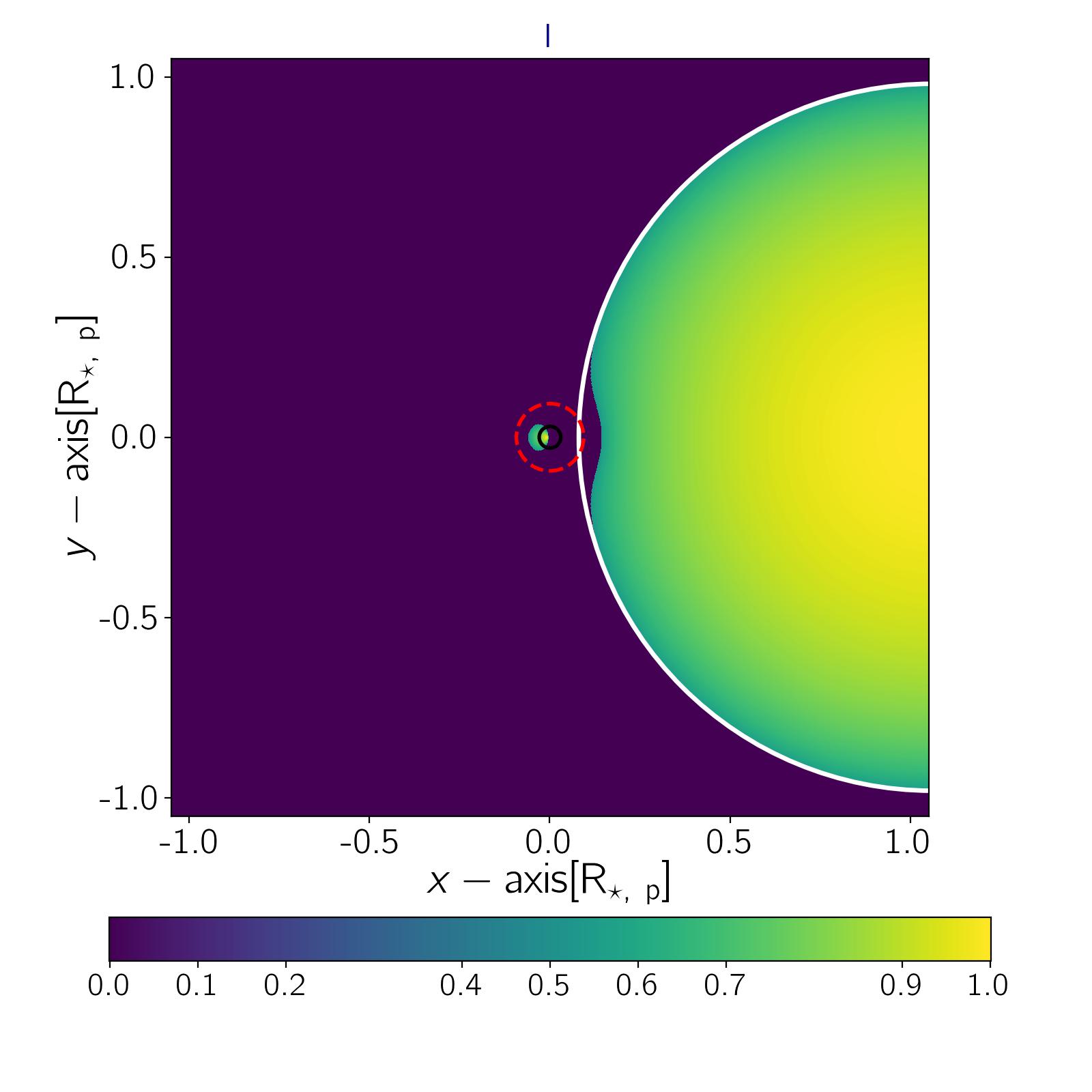}
\includegraphics[width=0.32\textwidth]{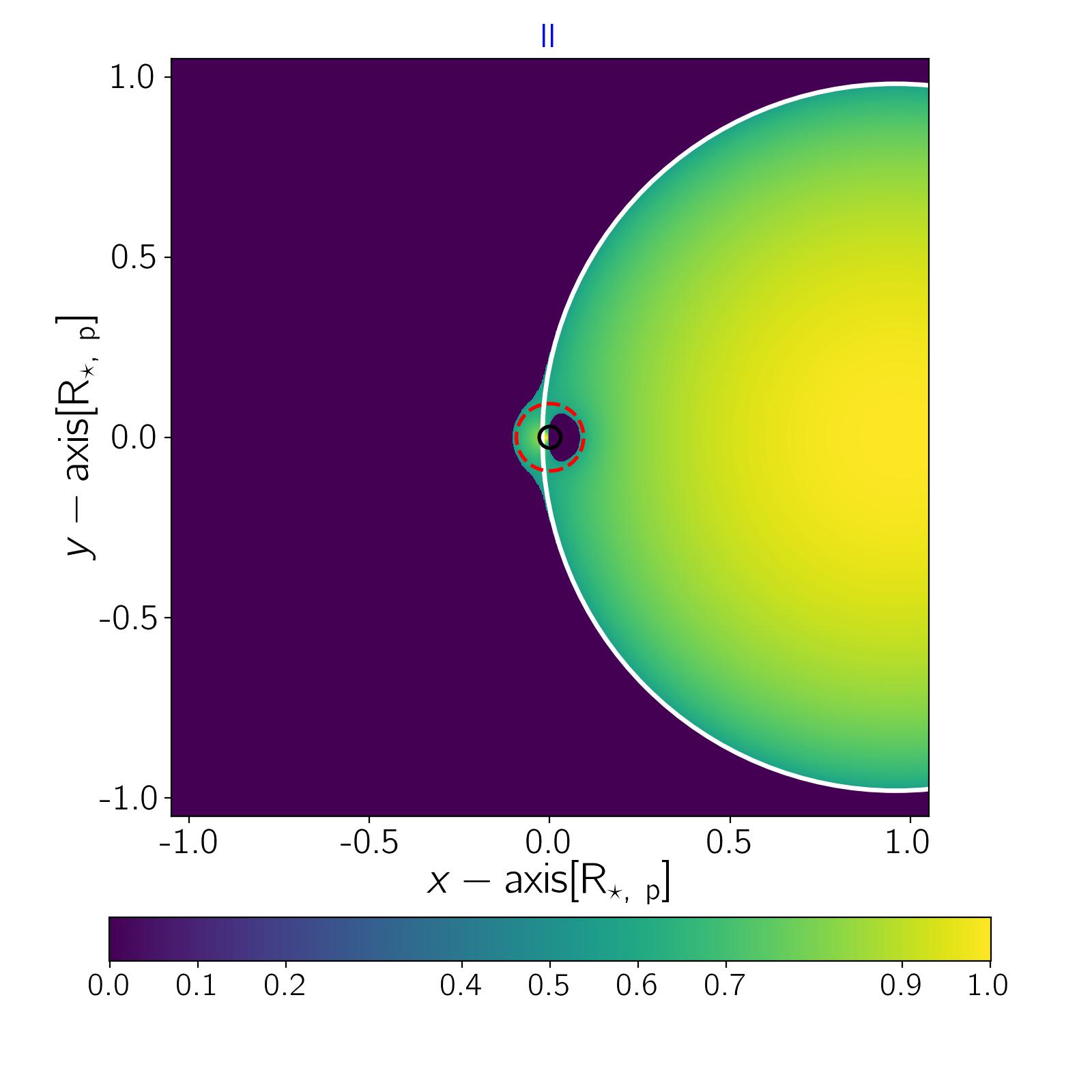}
\includegraphics[width=0.32\textwidth]{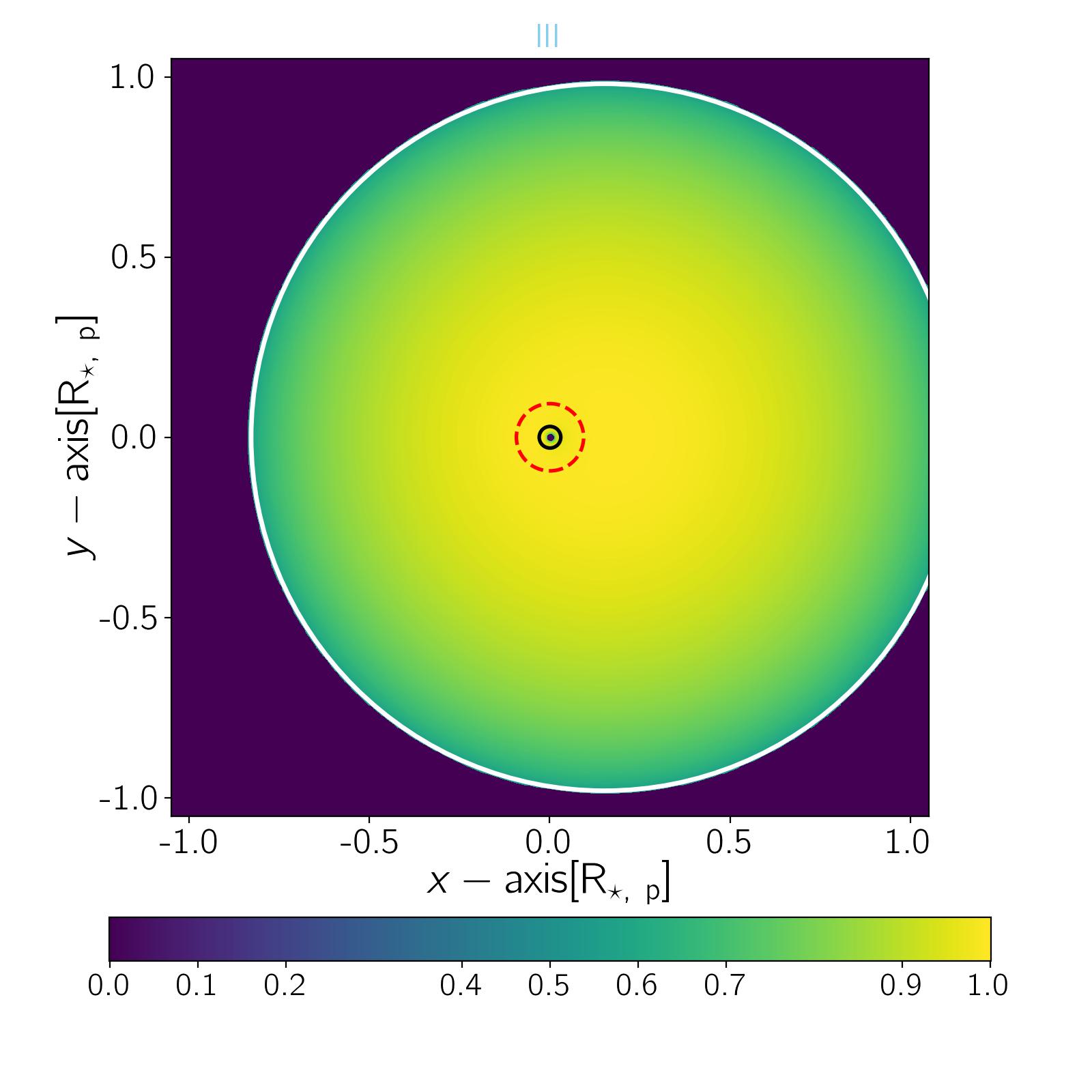}
\includegraphics[width=0.32\textwidth]{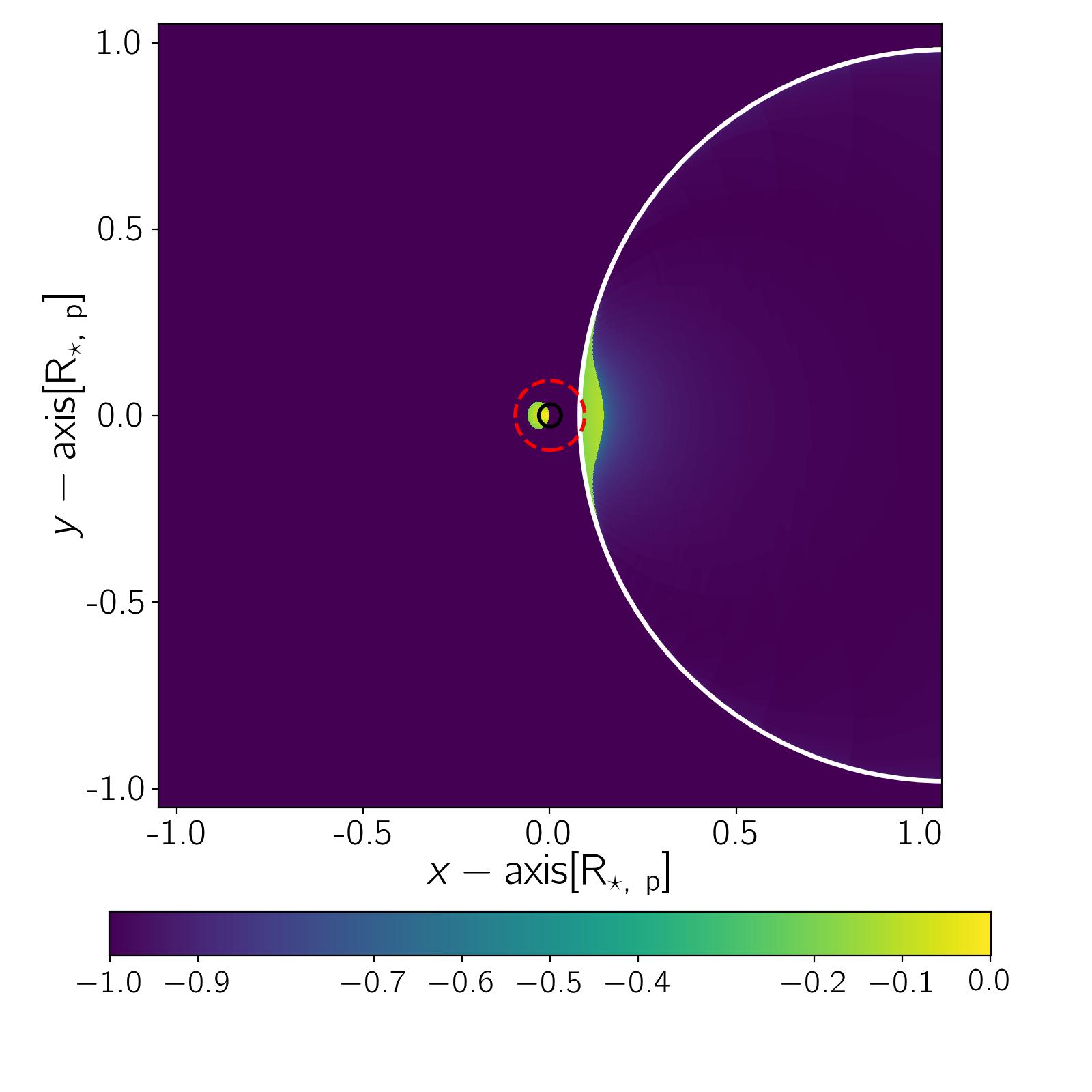}
\includegraphics[width=0.32\textwidth]{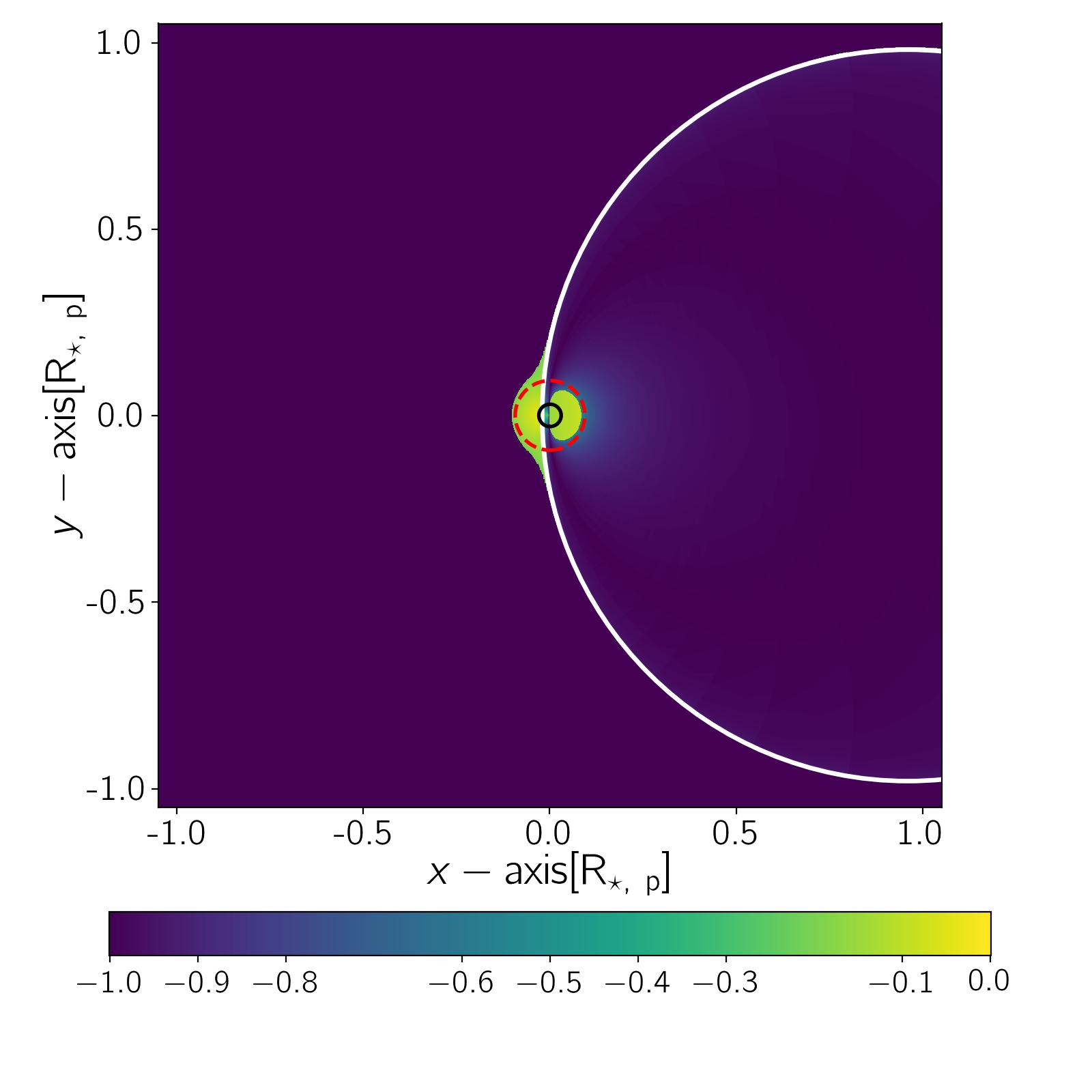}
\includegraphics[width=0.32\textwidth]{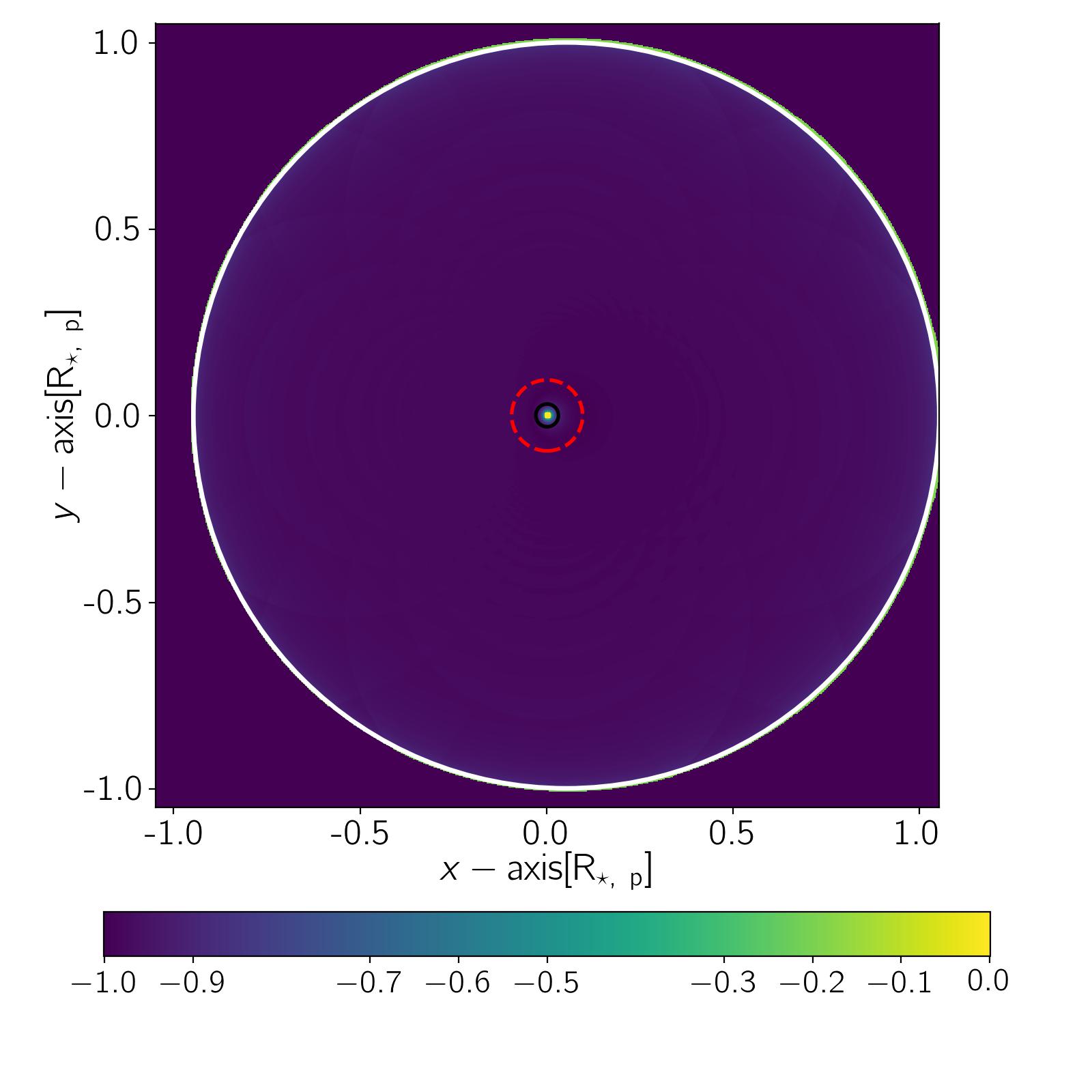}
\caption{Three top panels show the images formed during the self-lensing of a main-sequence star by its WD companion at three different times. Three bottom panels represent their corresponding residual maps between the projected brightnesses of images and the source star itself. The white, black, and red (dashed) circles display the edges of the source star, the lens object, and the Einstein radius. The axes are normalized to the source radius projected on the lens plane $R_{\star, \rm p}$. Two animations from  \href{https://iutbox.iut.ac.ir/index.php/s/mgoqo9YikR2W2se}{images maps}, and \href{https://iutbox.iut.ac.ir/index.php/s/BQ2CC3LfqBbgSpL}{residual maps} while the main-sequence star is passing behind the compact object versus time are available.}\label{fig1}
\end{figure*}

To simulate a self-lensing signal, we assume that the orbital plane of the binary system is edge-on as seen by the observer, i.e., the inclination angle is small. For an edge-on orbit, when the source star is passing behind the compact object its light is crossing the gravitational field of the compact object and is bent. By comparing the angular positions of the image ($\vartheta$) and the source star ($\beta$) in the lens plane, one can reach the raw form of the lens equation (without any approximation): 
\begin{eqnarray}
\tan\beta=\tan\vartheta-\big[\tan(\alpha-\vartheta) +\tan \vartheta  \big]\frac{D_{\rm ls}}{D_{\rm s}},  
\label{lenseq}
\end{eqnarray}  
where, $D_{\rm ls}=D_{\rm s}-D_{\rm l}$, and $D_{\rm s}$, and $D_{\rm l}$ are the source and lens distances from the observer. In the simulation, $D_{\rm l}$ which is the distance of the compact object from the observer is fixed. In our formalism, $D_{\rm s}=D_{\rm l}-x_{\rm o}$. We note that $\alpha\simeq\frac{4~G~M_{\rm{WD}}}{c^2~b}$ is the deflection angle due to the lensing effect, where $G$ is the gravitational constant, $c$ is the light speed, $b=\tan \vartheta~D_{\rm l}$ is the so-called impact parameter. 

Here, we calculate the magnification factor by solving the lens equation, \ref{lenseq}, using the inverse-ray-shooting (IRS) method \citep{1986AAKayser,1987AASchneiderw,1998LRRWambsganss,2010MNRAsajadian}. At each given time, we also calculate the fraction of the image's area that is covered by the lens disk, to evaluate the finite-lens effect.
\begin{figure*}
 \centering
\includegraphics[width=0.49\textwidth]{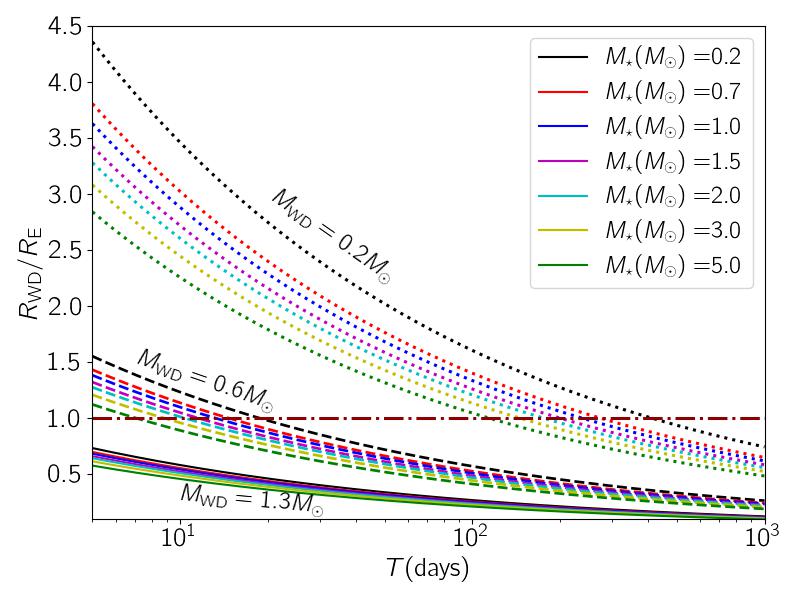}
\includegraphics[width=0.49\textwidth]{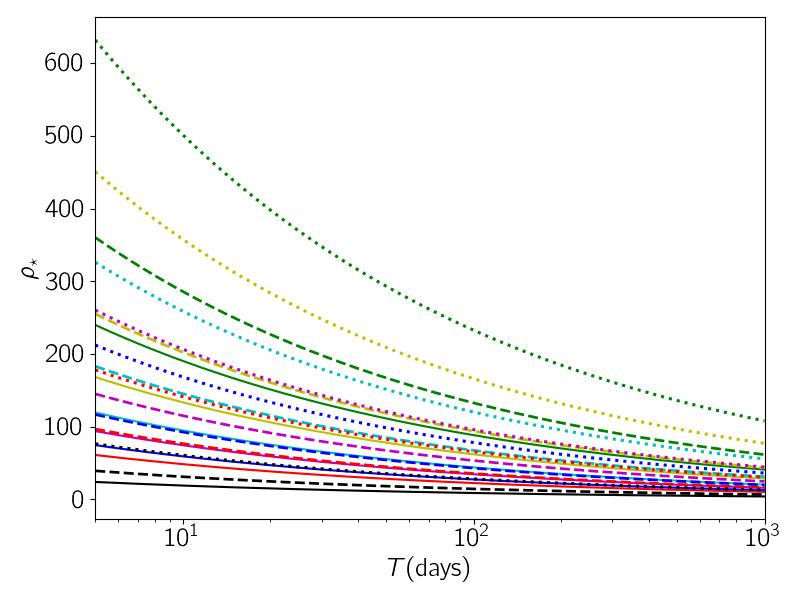}
\caption{These two plots represent $R_{\rm{WD}}/R_{\rm E}$ (left panel) and $\rho_{\star}$ (right panel) versus the orbital period $T(\rm{days})$ for different detached binary systems including WDs and main-sequence stars. Here, for detached binary systems three different values for the WD' mass $M_{\rm{WD}}(M_{\sun})=0.2,~0.6,~1.3$ (different line styles), and seven types of main-sequence stars with different masses (different colours of curves) are considered. The horizontal dark-red line in the left panel shows $R_{\rm{WD}}=R_{\rm E}$.}\label{RcNorm}
\end{figure*}

\noindent In our IRS calculations, the size of the lens plane is $5\rho_{\star} \times 5\rho_{\star}$, and we divide it into $2500\times 2500$ grids. The position of each grid, $(x_{\rm i},~y_{\rm i})$, determines the lensing impact parameter $b=R_{\rm E}\sqrt{x_{\rm i}^{2}+y_{\rm i}^{2}}$, and accordingly $\tan \vartheta=b/D_{\rm l}$. Using Equation \ref{lenseq}, one can indicate the initial line of sight of that light (if it was not bent), i.e., $\beta$, and its components, i.e. $(x_{\beta},~y_{\beta})= \eta \times (x_{\rm i},~y_{\rm i})\big/b$, where $\eta= D_{\rm l} \tan \beta$. Here, we use the fact that the gravitational lens does not change the azimuthal angle of the light. If $(x_{\beta},~y_{\beta})$ is one point on the source disk (which is a circle with the radius $\rho_{\star}$, and the center $(y_{\rm o}/R_{\rm E},~z_{\rm o}/R_{\rm E})$), the point $(x_{\rm i},~y_{\rm i})$ is over the images' disk. Also, if $b\leq R_{\rm{WD}}$, we do not receive the light of that part of the images as it is obscured by the lens's disk.

In this formalism, we also consider the linear limb-darkening profile for the surface brightness of the source star, as $I=I_{0} \big(1-\Gamma [1-\mu]\big)$, where $I_{0}$ is the brightness at the source's center, $\Gamma$ is the linear limb-darkening coefficient, $\mu=\sqrt{1-R^{2}/R_{\star}^{2}}$, and $R$ is the radial distance over the source disk. Based on this explained formalism, we generate the images in self-lensing events and discuss their properties in the next subsection.

\subsection{Finite-Lens Effect  vis  Self-Lensing}\label{sec3_2}
As explained in Section \ref{sec2}, in a self-lensing event the source star and lens object are very close to each other, i.e., $\big|x_{\rm o}\big|\ll D_{\rm l}$. Hence, in these events the Einstein radius which is $R_{\rm E}=\sqrt{\frac{4~G~M_{\rm{WD}}}{c^{2}}\frac{D_{\rm l} |x_{\rm o}| }{D_{\rm l}-x_{\rm o}}}$ is small ($\sim 0.02-0.05 R_{\sun}$) in comparison with the Einstein radius of a common microlensing event toward the Galactic bulge (i.e., $\sim 1-2 \rm{AU}$). Therefore, in self-lensing events (i) finite-source size is large $\rho_{\star} \gtrsim1$, and (ii) finite-lens size could be significant. 

We simulate the images of a source star lensed by its compact companion while the source star is passing behind it versus time using the IRS method. In the three top panels of Figure \ref{fig1}, we show the images in three different positions of the source star projected on the lens plane. In these figures, the white, black, and red circles specify the source star, lens object, and the Einstein ring, respectively. Accordingly, lensing effects on the source stars are barely realizable. Hence in their bottom panels, we show the residual maps (the absolute values of difference between the images' maps and the source maps in the logarithmic scale). To make these maps we use the following parameters: $M_{\star}=0.2M_{\sun}$, $R_{\star}=0.3R_{\sun}$, $M_{\rm{WD}}=1.3M_{\sun}$, $R_{\rm{WD}}=0.002R_{\sun}$, $T=60$ days, $i=0^{\circ}$, $\theta=0^{\circ}$, $\epsilon=0$, $D_{\rm l}=1$ kpc, and $\Gamma=0.5$.

Accordingly, in lensing events due to a single lens object from a large source star, if the lens is out of the source disk ($u>\rho_{\star}$) two images are formed. One of them is inside the Einstein radius and very small, and another one is similar to the source star with some deformation at the source's edge. When the lens is over the projected source disk, inside the Einstein ring (the red dashed circles) there are (i) an empty hole (de-magnification) so that its size decreases while the projected distance between the source centre and the compact object is decreasing, and (ii) a lensed part (in the opposite side of that empty hole). Therefore and according to the three bottom panels, most of the variations during self-lensing happen inside the Einstein radius. If the lens radius is much smaller than the Einstein radius $R_{\rm{WD}}\ll R_{\rm E}$, the finite-lens effect is ignorable. But if the lens radius is comparable with the Einstein radius, it will change the overall magnification factor. There is an empty hole in the Einstein radius, and hence the occultation amount depends on the radius of that hole in addition to the $R_{\rm WD}$, and $R_{\rm E}$. We study this point in the next section.  
\begin{figure*}
\centering
\subfigure[]{\includegraphics[width=0.45\textwidth]{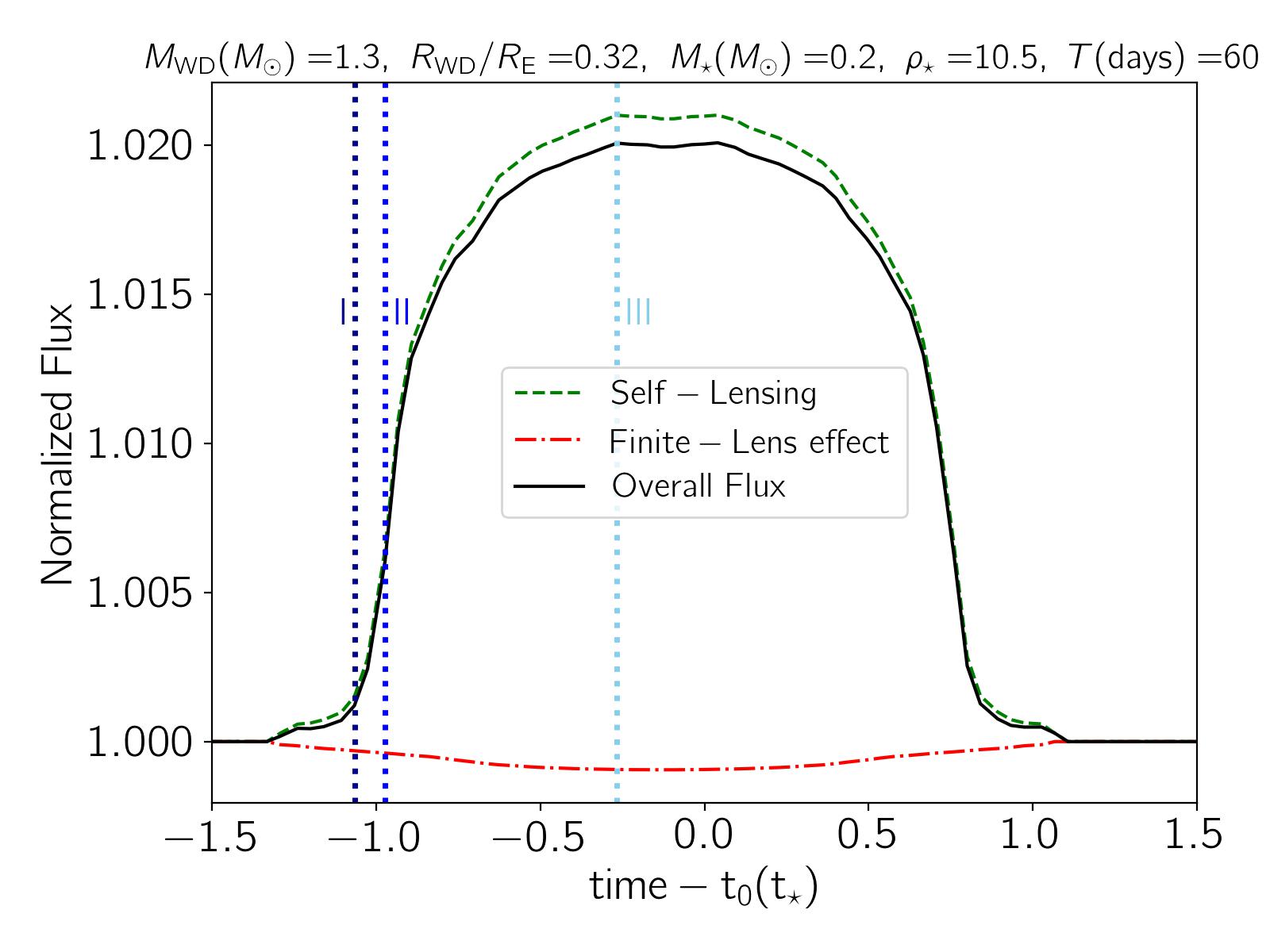}\label{lighta}}
\subfigure[]{\includegraphics[width=0.45\textwidth]{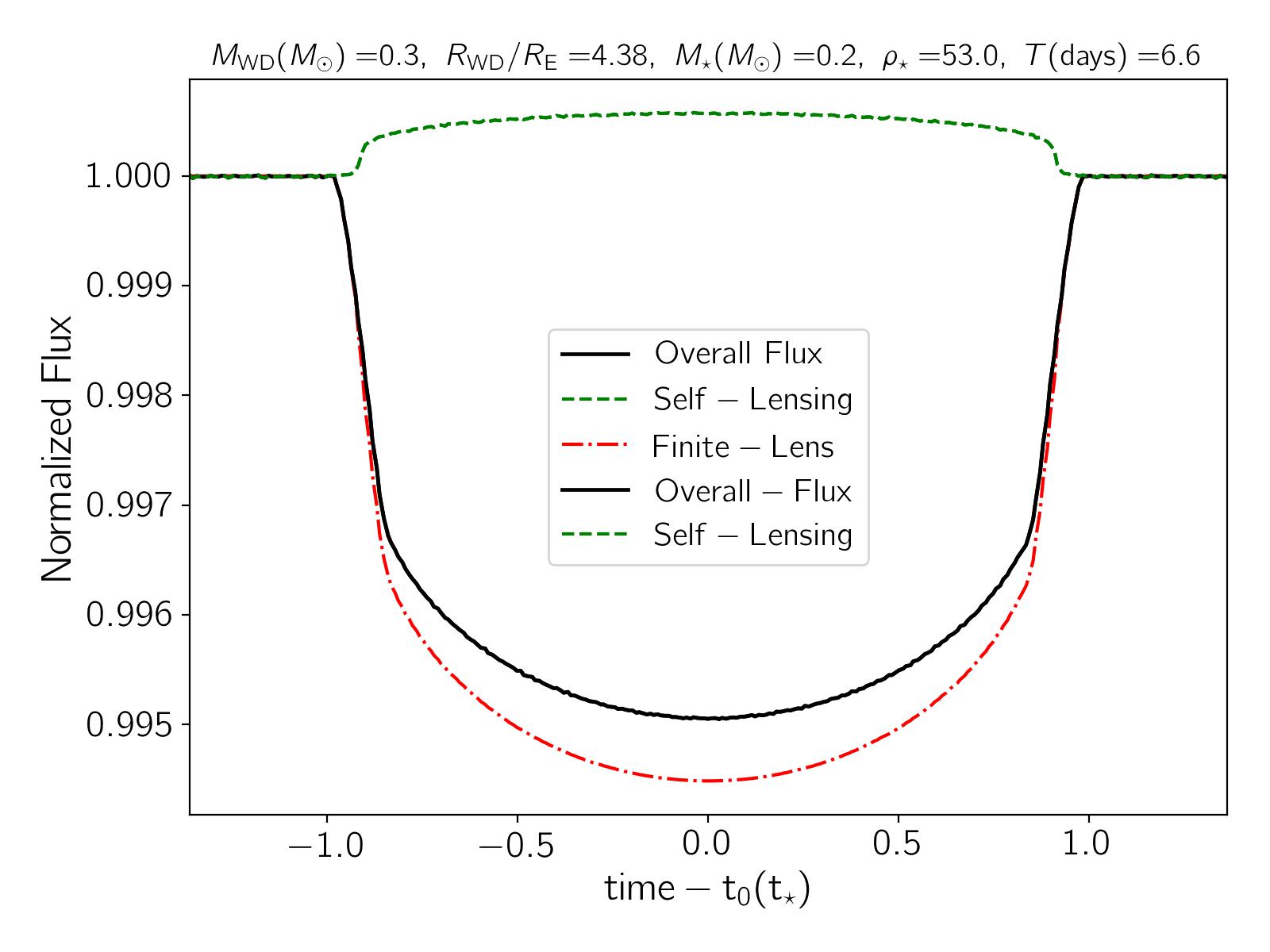}\label{lightb}}
\subfigure[]{\includegraphics[width=0.45\textwidth]{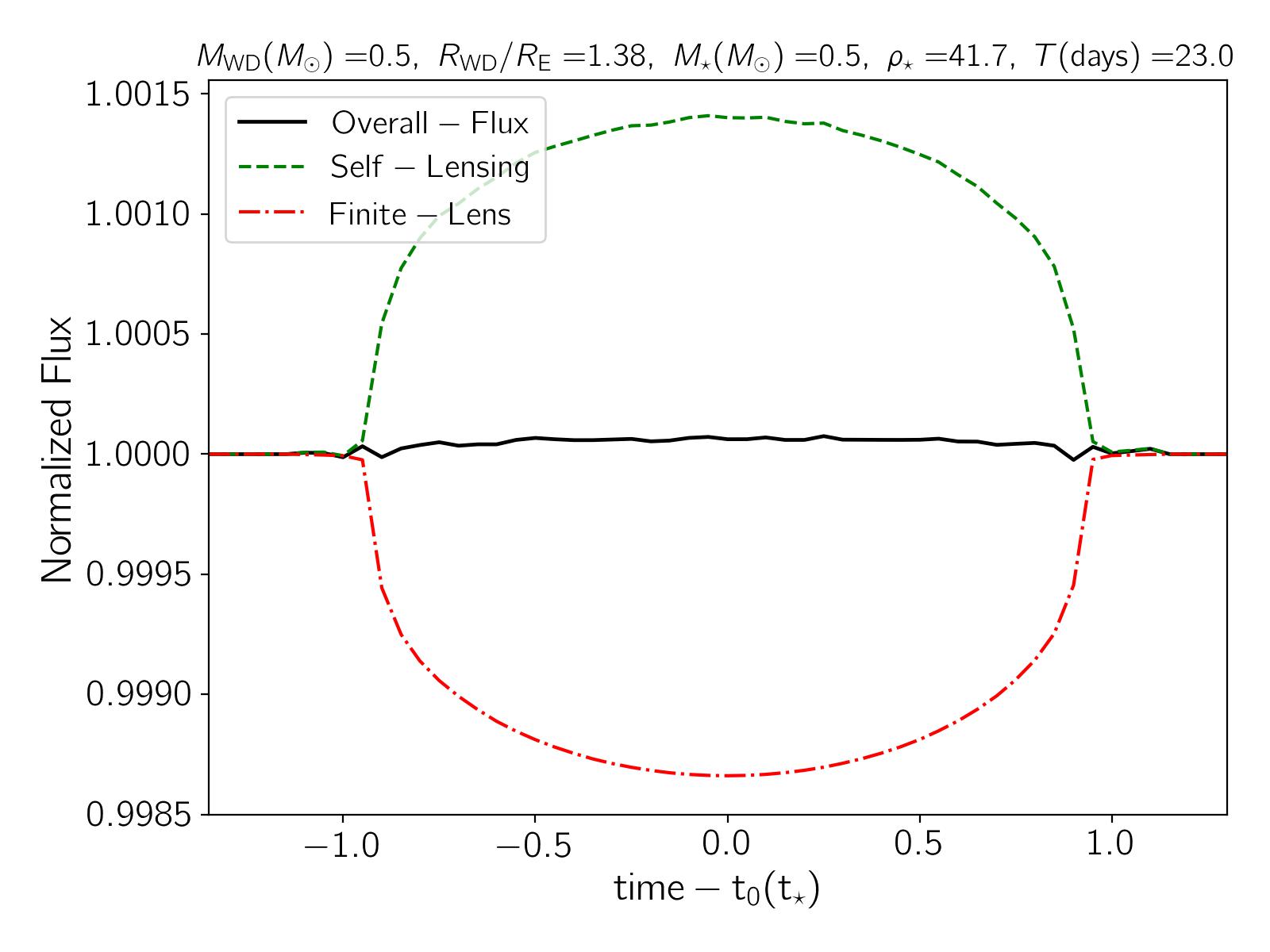}\label{lightc}}
\subfigure[]{\includegraphics[width=0.45\textwidth]{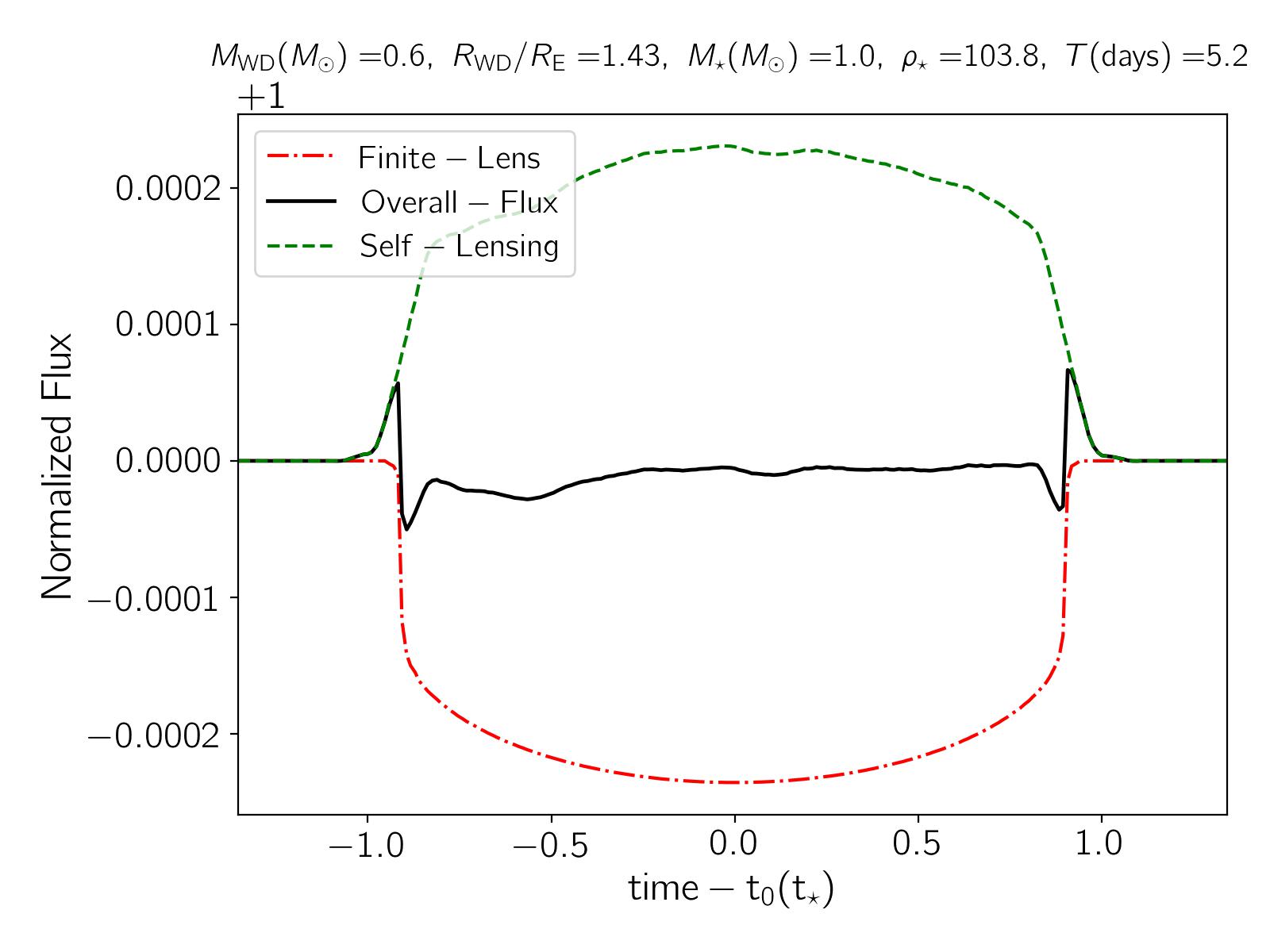}\label{lightd}}
\subfigure[]{\includegraphics[width=0.45\textwidth]{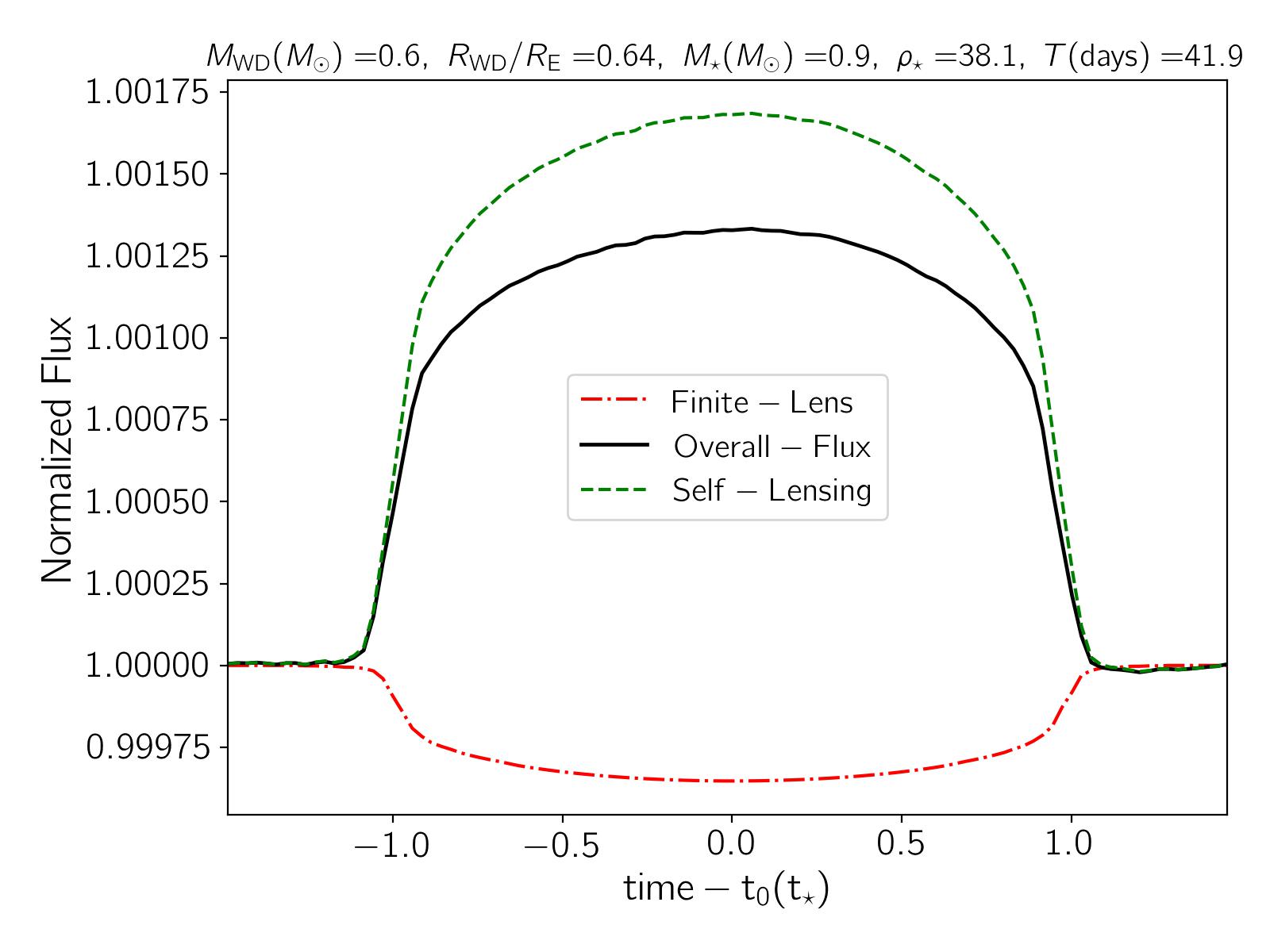}\label{lighte}}
\subfigure[]{\includegraphics[width=0.45\textwidth]{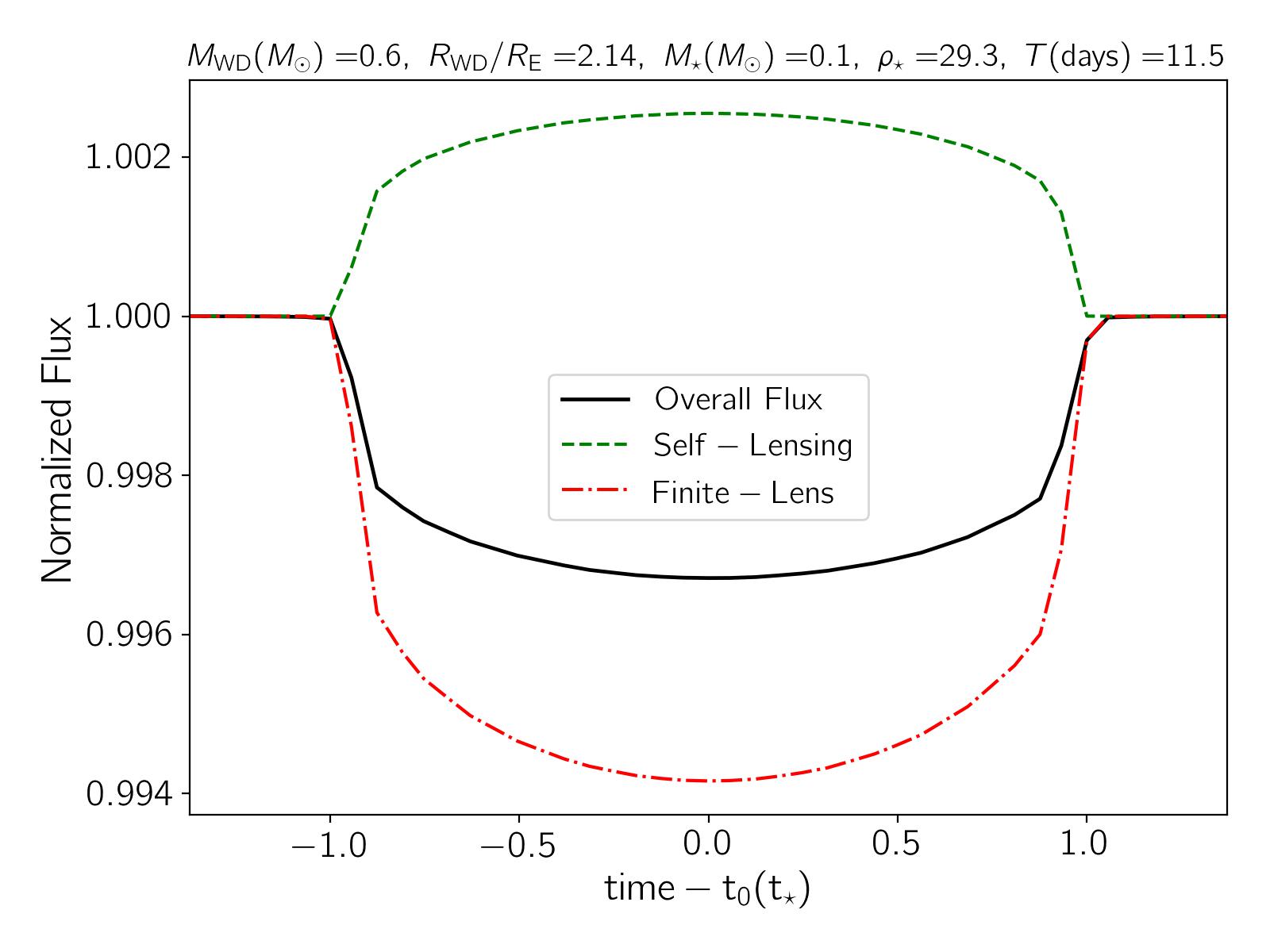}\label{lightf}}
\caption{Examples of stellar light curves due to WDMS binary systems while the source stars are passing behind the compact objects. In these panels, self-lensing curves, obscuration curves due to finite-lens effects, and the overall light curves due to both self-lensing and finite-lens effects are shown by green dashed, red dot-dashed, and black solid curves, respectively. The relevant parameters are given at the top of panels. The first panel shows the light curve due to the WDMS binary system whose lensing-induced images are represented in Figure \ref{fig1}. }\label{Light2}
\end{figure*}
  
We investigate in which kinds of WDMS binary systems the lens radius has a similar order of magnitude with the Einstein radius or even is larger than it, $R_{\rm{WD}}\gtrsim R_{\rm E}$. For WDs, there is a known decreasing relation between their mass and radius as \citep[e.g., see, ][]{2020Ambrosino}: 
\begin{eqnarray}
R_{\rm{WD}}(R_{\sun})=0.01(M_{\sun}\big/M_{\rm{WD}})^{1/3}. 
\label{wdrm}
\end{eqnarray} 
For instance, a typical WD with $M_{\rm{WD}}\simeq 0.6 M_{\sun}$ has the radius $R_{\rm{WD}} \simeq0.012 R_{\sun}$. Using Equation \ref{wdrm} $R_{\rm{WD}}/R_{\rm E}$ is given by: 
\begin{eqnarray}
\frac{R_{\rm{WD}}}{R_{\rm E}}\simeq 0.36~[T(\rm{day})]^{2/3} \big(\frac{M_{\rm{WD}}}{M_{\sun}}\big)^{5/3} \Big[\frac{M_{\rm{WD}}}{M_{\sun}}+\frac{M_{\star}}{M_{\sun}}\Big]^{1/3}.  
\label{Rwdre}
\end{eqnarray}
By considering seven types of main-sequence stars and three values for WDs' mass, we calculate $R_{\rm{WD}}/R_{\rm E}$ as shown in the left panel of Figure \ref{RcNorm} versus the orbital period $T(\rm{days})$. In Equation \ref{Rwdre}, we calculate the semi-major axis (for deriving the Einstein radius) using the Kepler's Third law. 

According to this plot, for most of WDs and source stars with $M_{\rm WD}\gtrsim 0.2 M_{\sun}$, and $M_{\star}\in [0.08,~1.2] M_{\sun}$ if their orbital periods are $T\gtrsim 300$ days (where $R_{\rm WD}\lesssim R_{\rm E}$ in the left panel of Figure \ref{RcNorm}), the magnification due to self-lensing dominates the occultation owing to the finite-lens effect. Most massive WDs with $M_{\rm WD}\gtrsim1.3 M_{\odot}$ (which is close to the Chandrasekhar limit) have very small radii. For these WDs, even by considering a wide range of orbital periods $T \in [5,~1000]$ days and different values of $M_{\star}$, we do not expect significant finite-lens effect, because they have $R_{\rm{WD}}\lesssim 0.6R_{\rm E}$. For common WDs with $M_{\rm WD}\simeq 0.6 M_{\sun}$ if the orbital period is less than $\sim 20$ days, their binary systems have $R_{\rm WD}\gtrsim R_{\rm E}$, i.e., the occultation effect is considerable.

To evaluate the self-lensing signals in these binary systems, in the right panel of Figure \ref{RcNorm} we display $\rho_{\star}$ for the mentioned binary systems versus the orbital period. For WDMS binary systems we can estimate $\rho_{\star}$ as
\begin{eqnarray}
\rho_{\star}\simeq 23.4\frac{R_{\star}(R_{\sun})}{\sqrt{M_{\rm{WD}}(M_{\sun})}}\Big[\frac{1}{\sqrt{a (\rm{au})}} -2.4\times 10^{-9}\frac{\sqrt{a(\rm{au})}}{D_{\rm l}(\rm{kp})}\Big].
\end{eqnarray}
Since the self-lensing signal can be estimated by $1+2 \rho_{\star}^{-2}$ \citep[see, e.g., ][]{1973AAMaeder,1996Gould,2003ApJEric}, the smaller the projected and normalized source radii, the larger the self-lensing signals. 

\noindent According to two panels of Figure \ref{RcNorm}, self-lensing signals dominate finite-lens effects when $R_{\rm E}$ is large. Hence, by either increasing the mass of WD or the orbital period or both of them, the self-lensing signal is enhanced and the finite-lens effect is reduced. The highest self-lensing signal happens for most massive WDs ($M_{\rm{WD}}\sim 1.3 M_{\odot}$) that have dwarf companions in wide orbits. For that reason, all five discovered self-lensing events have $M_{\rm WD}\ge 0.5 M_{\odot}$, and $T(\rm{days})\ge88$ \citep{KruseAgol2014,2018AJKawahara,2019ApJLMasuda,2024Yamaguchi}. For the self lensing target KIC 8145411, although the initial estimation for the WD mass was $0.2 M_{\odot}$ which was in conflict with the standard binary evolutionary path \citep{2019ApJLMasuda}, the future high-resolution photometry data revealed the existence of a third component and indicated the WD mass should be $0.5 M_{\odot}$ \citep{2024Yamaguchi}.

We note that both $M_{\rm{WD}}$ and $a$ (which is $\propto T^{2/3}$) have the same effects on $R_{\rm E}$, so that increasing the first one from $0.2M_{\sun}$ to $1.3M_{\sun}$ or increasing the orbital period from $10$ to $200$ days both reduce $\rho_{\star}$ by a factor of $0.4$. Inversely, the occultation effect due to finite-lens size dominates the self-lensing signal when $R_{\rm E}$ is small which results in large $\rho_{\star}$, and a considerable occultation. We note that both effects (the magnification and occultation) enhance when the source radius reduces. In the next section, we study stellar light curves that contain self-lensing signals affected by the finite-lens effect.

\begin{figure*}
\centering
\includegraphics[width=0.32\textwidth]{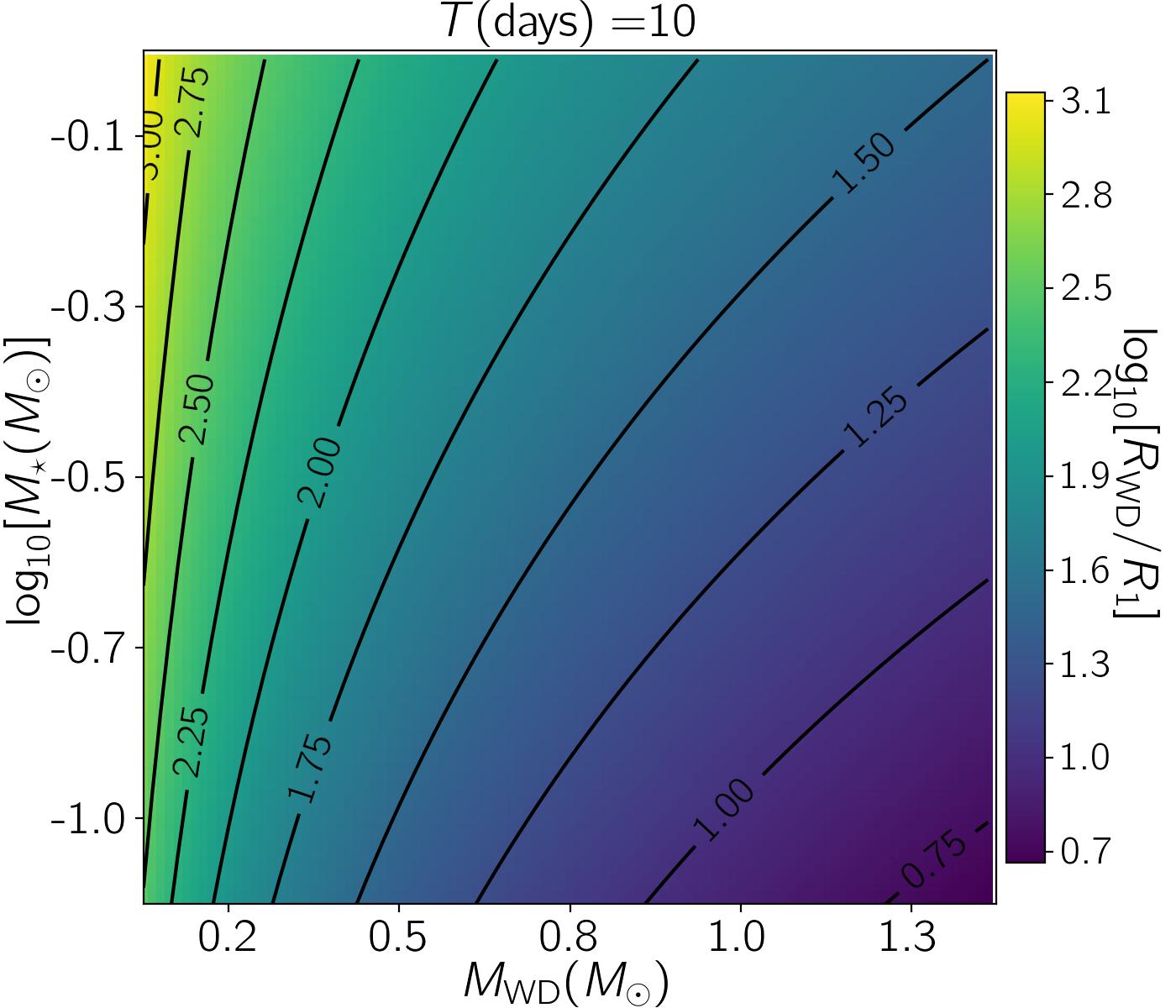}
\includegraphics[width=0.32\textwidth]{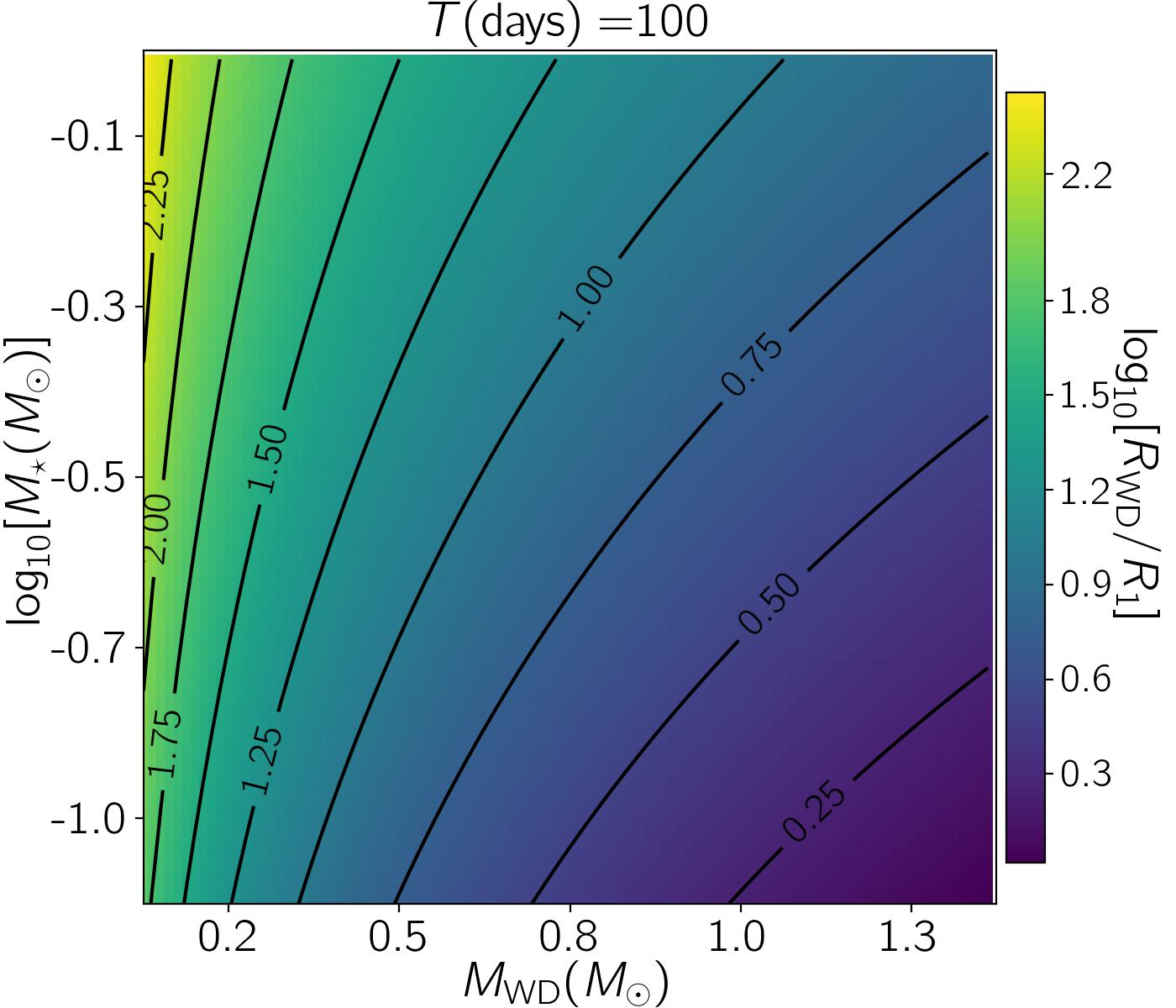}
\includegraphics[width=0.32\textwidth]{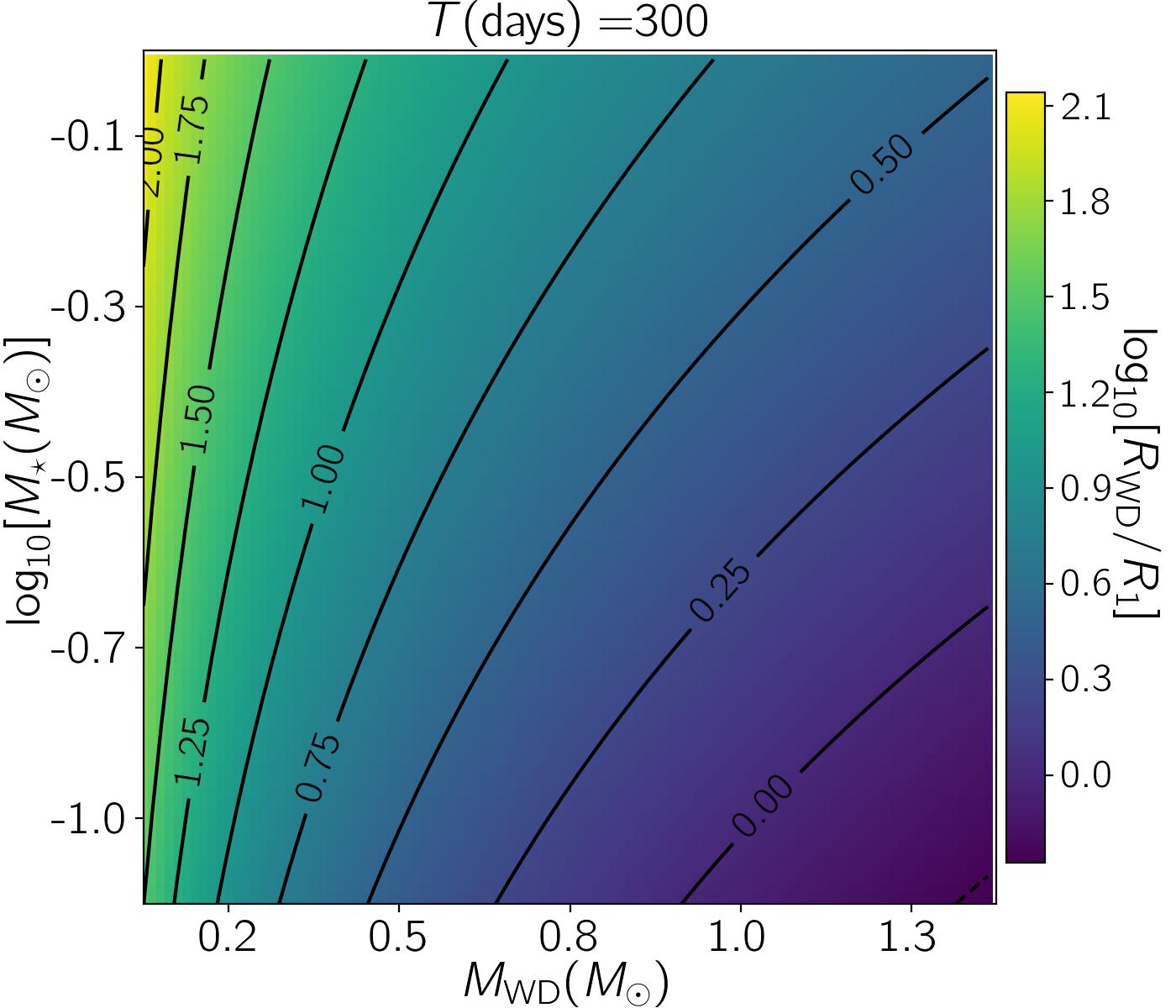}
\includegraphics[width=0.32\textwidth]{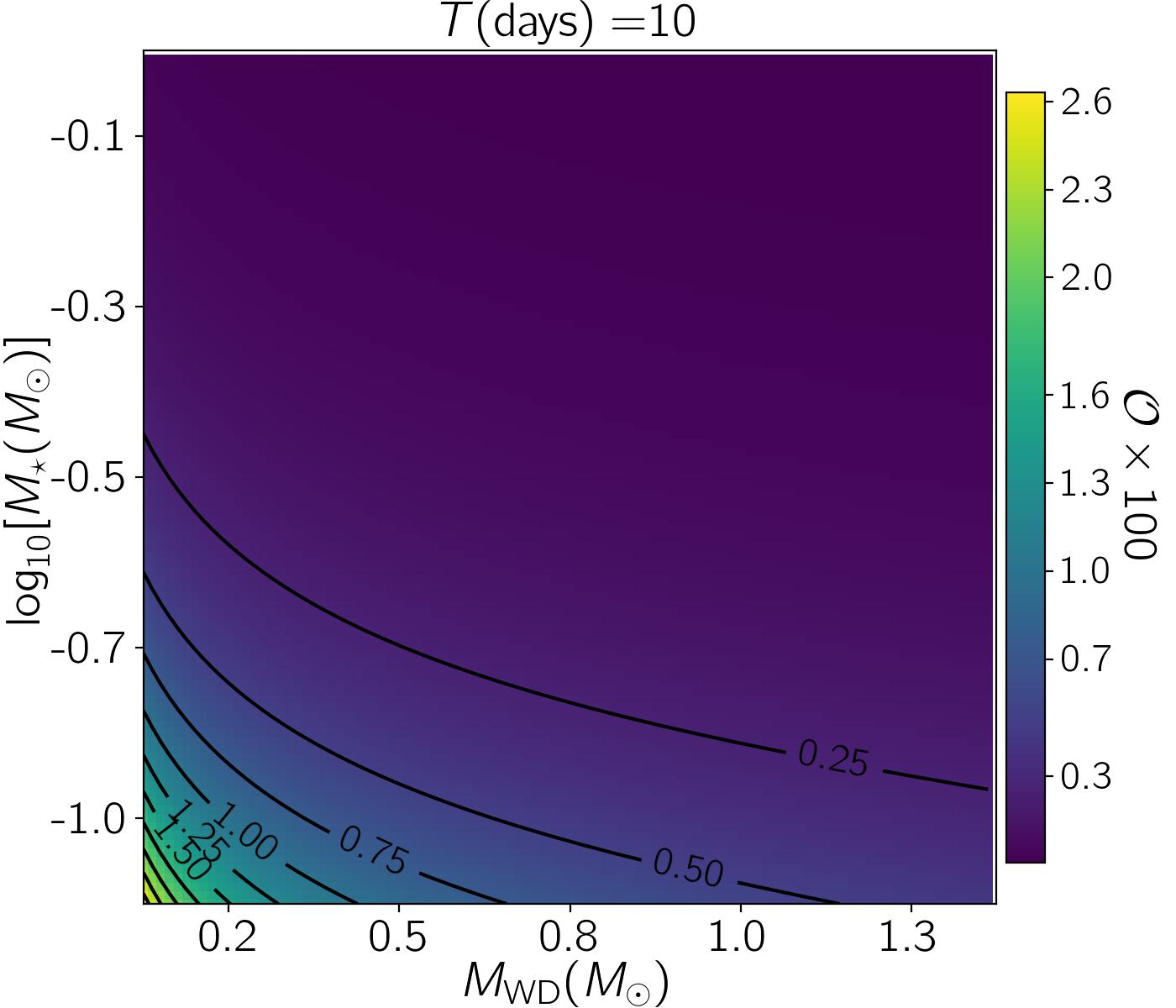}
\includegraphics[width=0.32\textwidth]{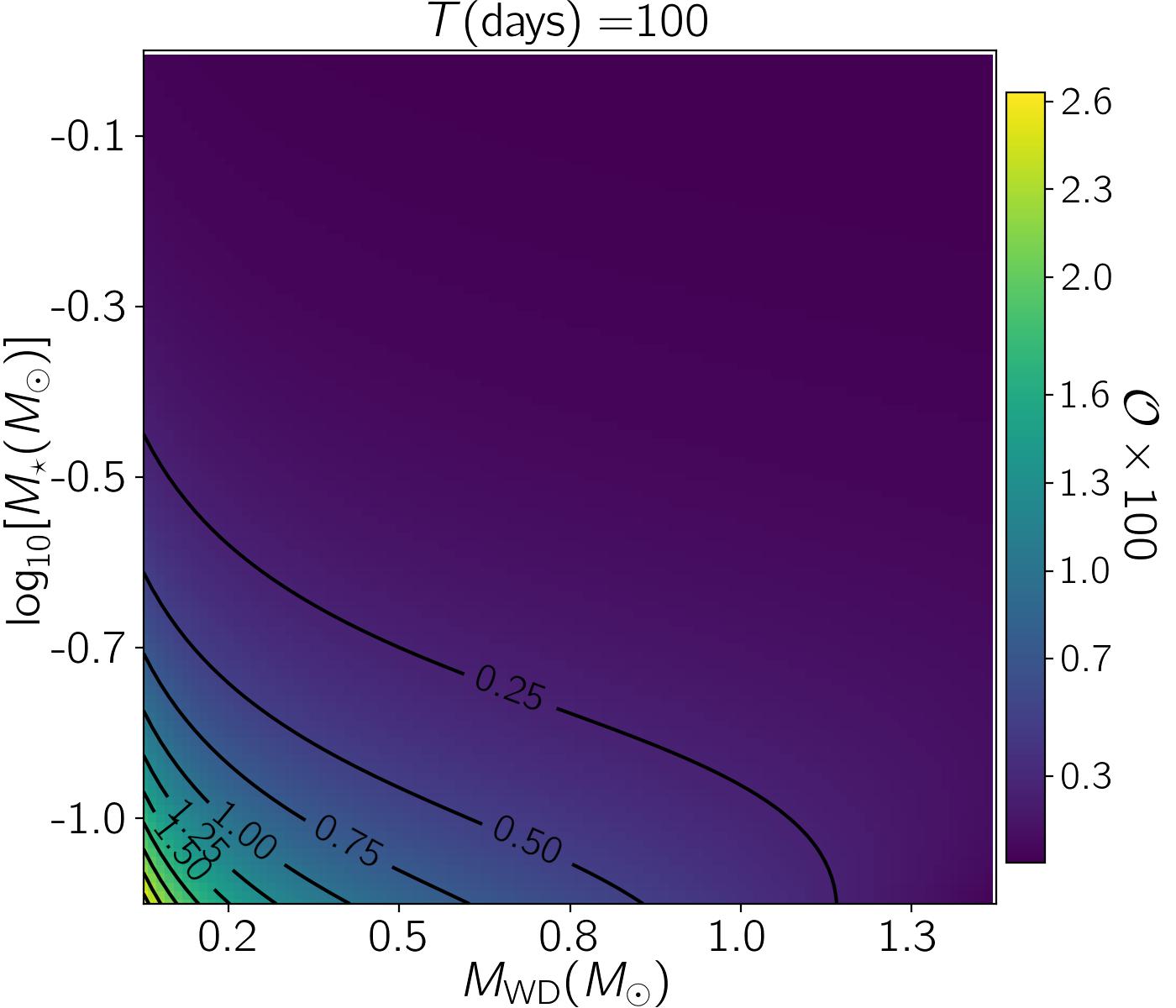}
\includegraphics[width=0.32\textwidth]{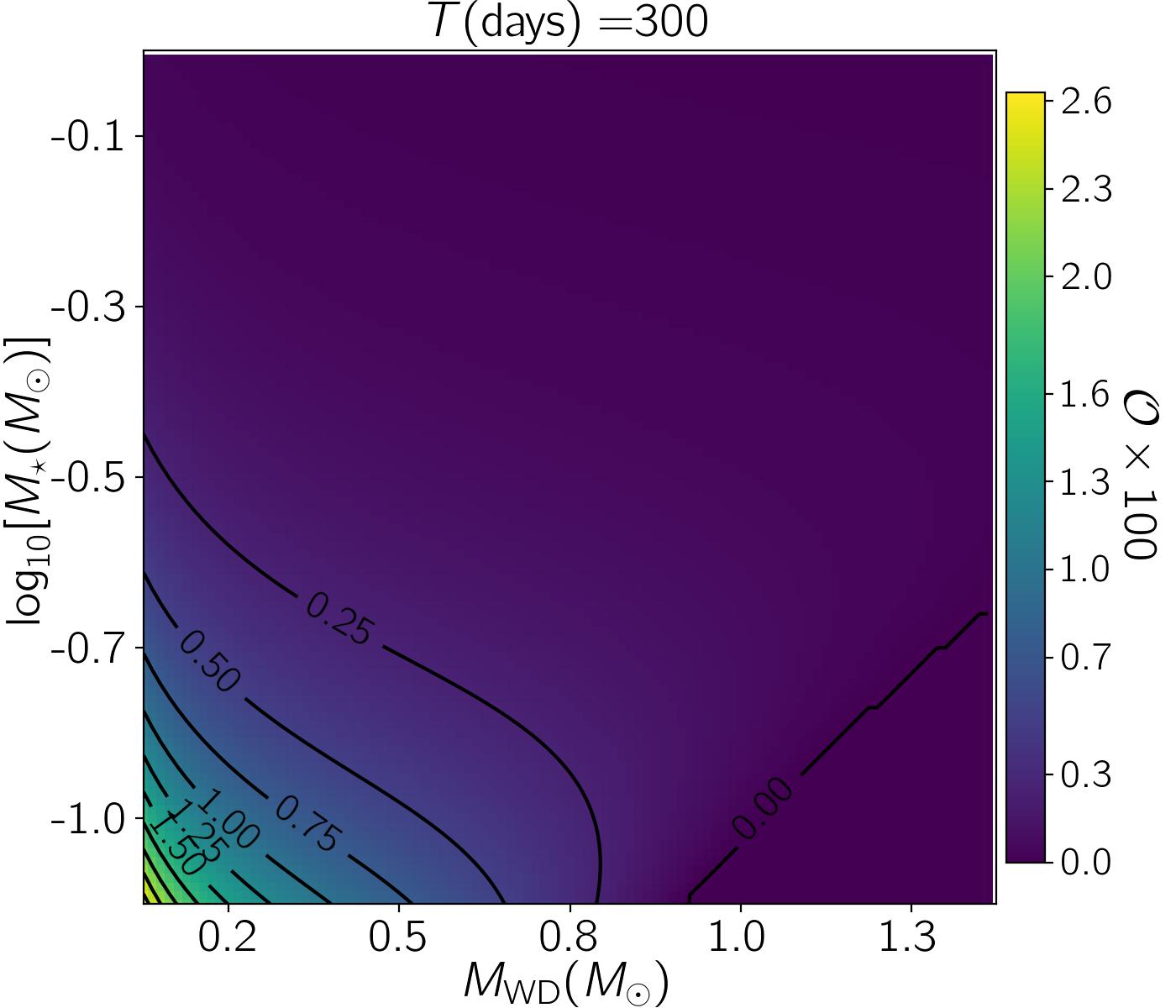}
\includegraphics[width=0.32\textwidth]{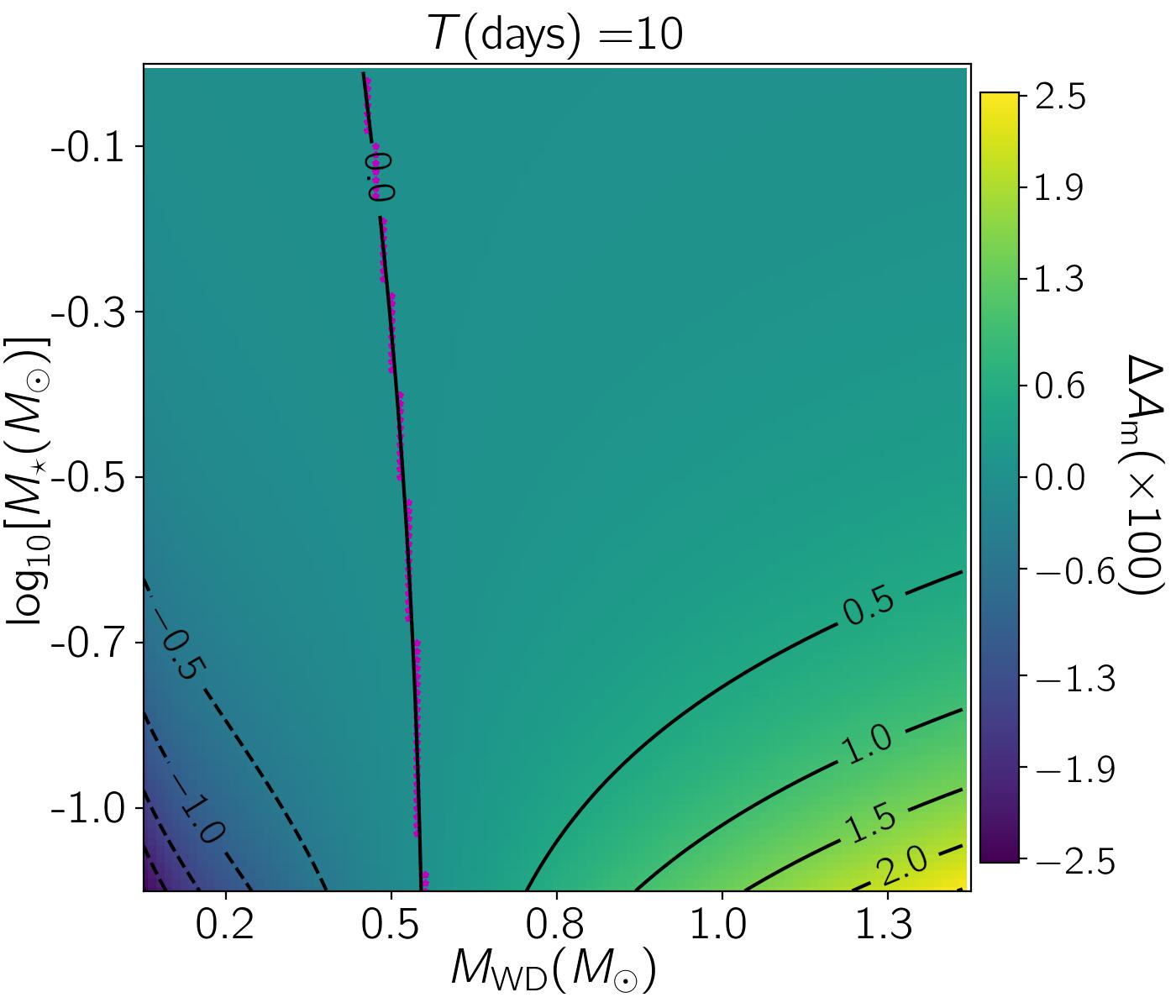}
\includegraphics[width=0.32\textwidth]{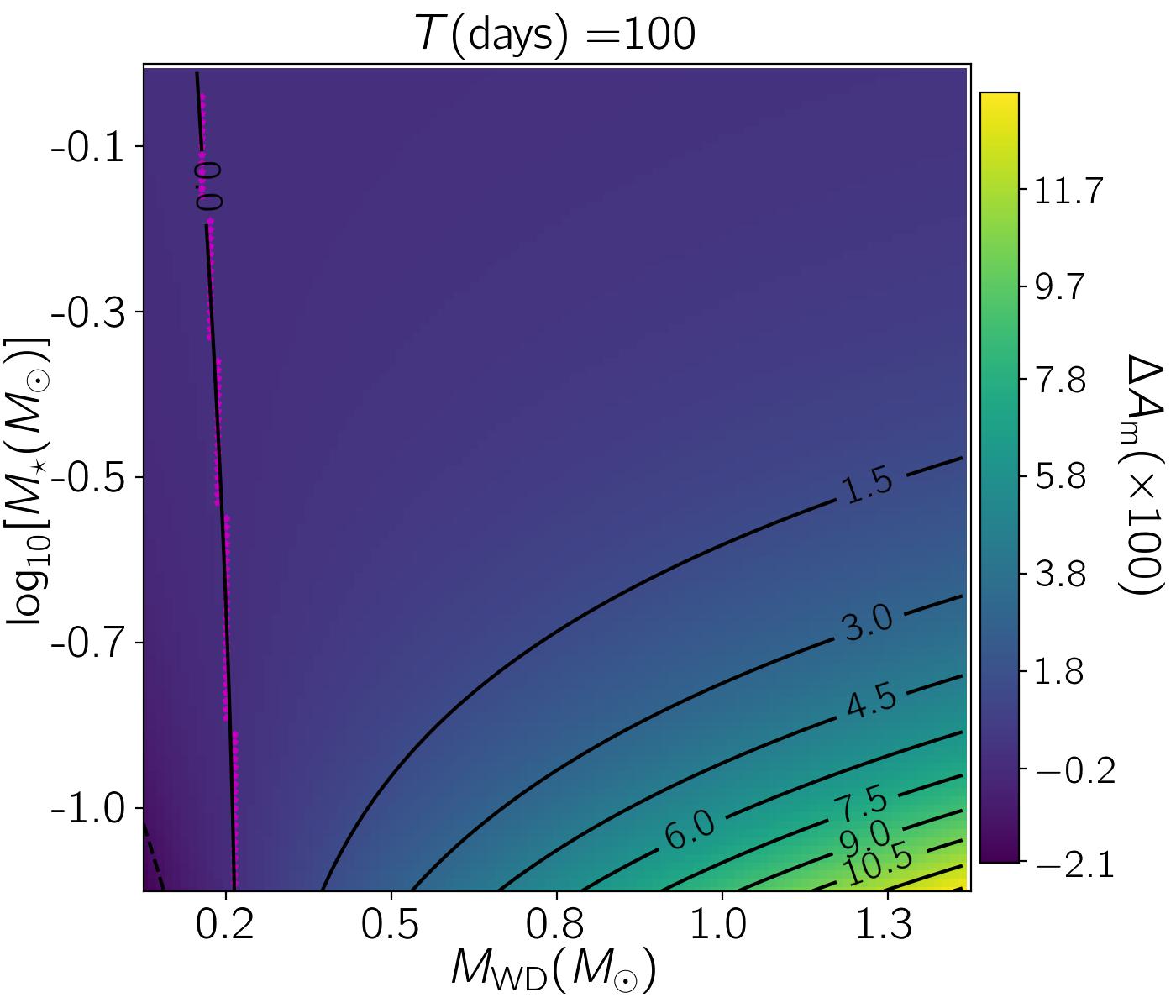}
\includegraphics[width=0.32\textwidth]{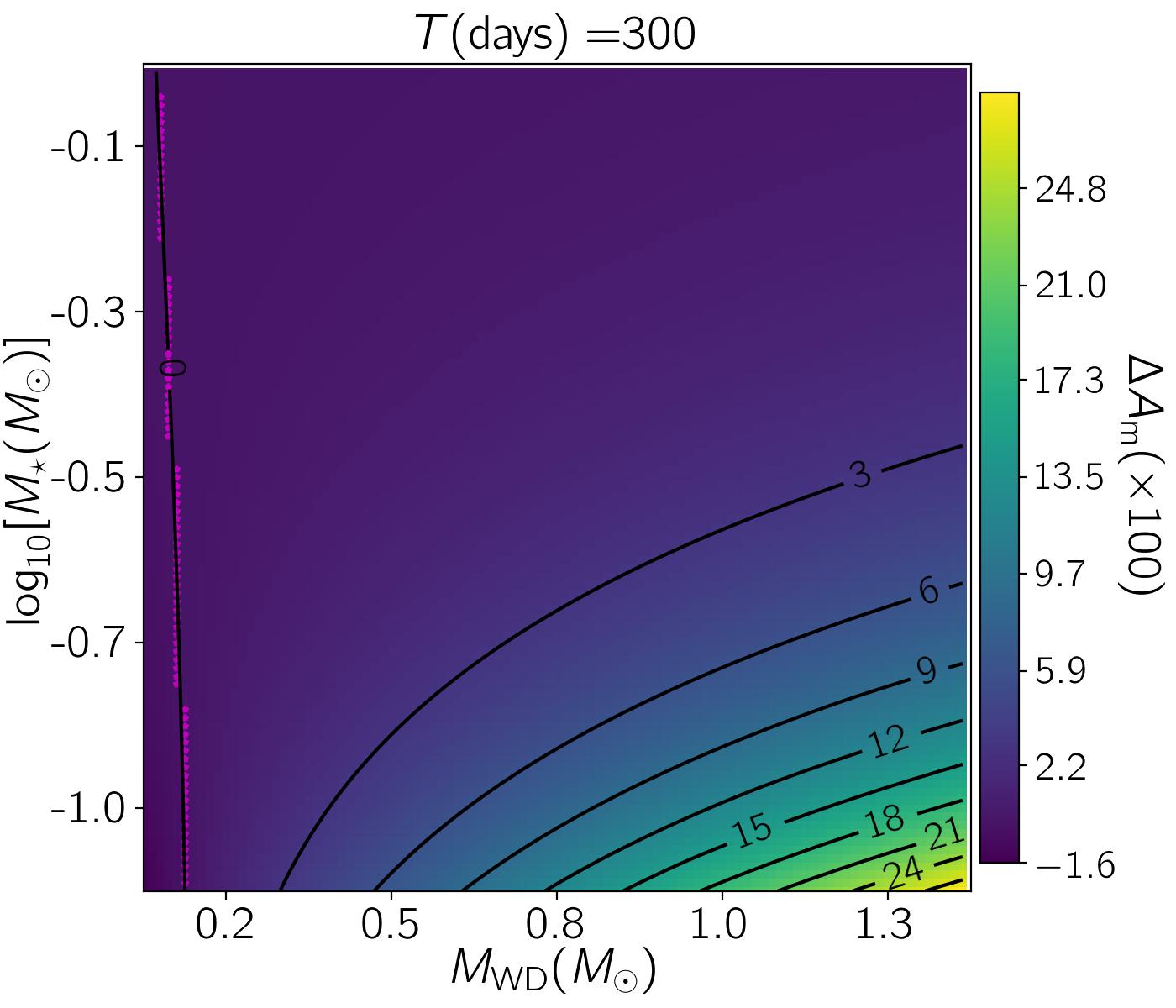}
\caption{For three orbital periods $T=10,~100,~300$ days (from left to right), the maps of $\log_{10}[R_{\rm{WD}}/R_{1}]$, $\mathcal{O}\times 100$, and $\Delta A_{\rm{m}}\times 100$ over the 2D space $M_{\rm{WD}}(M_{\sun})-\log_{10}[M_{\star}(M_{\sun})]$ are displayed from top to bottom. The black curves show the contour lines for each map. Three animations from the maps of \href{https://iutbox.iut.ac.ir/index.php/s/gmxLytwMJWN84ni}{$\log_{10}[R_{\rm{WD}}/R_{1}]$},  \href{https://iutbox.iut.ac.ir/index.php/s/QDJ8sfbbTdgiTKT}{$\mathcal{O}\times 100$}, and \href{https://iutbox.iut.ac.ir/index.php/s/tezLS7jKiHCzMiW}{$\Delta A_{\rm{m}}\times 100$} versus the orbital period are available.}\label{mapp}
\end{figure*}
\section{Self-Lensing/Occultation Light Curves}\label{sec4}
Using the IRS method we make a sample of stellar light curves for WDMS binary systems while the source stars are passing behind the compact companions. Some of these light curves are shown in Figure \ref{Light2}. In each panel, the normalized fluxes versus time due to the self-lensing signal, occultation effect, and overlapped self-lensing/occultation effects are shown by green dashed, red dot-dashed, and black solid curves, respectively. The $x$-axis is normalized to $t_{\star}$, which is the time of crossing the source radius. The key parameters to make each of them are mentioned at the top of each panel.

\noindent The first panel, \ref{lighta}, represents the light curve due to the WDMS binary system whose lensing-induced images (at three different times labeled with I, II, and III) are represented in top panels of Figure \ref{fig1}. In this light curve, the self-lensing signal is dominant. In the next panel, \ref{lightb}, the occultation effect is dominant. $R_{\rm E}$ values for these two WDMS binary systems are $0.029 R_{\sun}$ and $0.005 R_{\sun}$, respectively. 

If the enhancement due to the self-lensing and occultation effects have similar orders of magnitude, they cancel each other out. By simulating different stellar light curves from edge-on and detached WDMS binary systems, we find that when $R_{\rm{WD}}\simeq 1.4 R_{\rm E}$, and for any given value of $\rho_{\star}$ in their stellar light curves there are neither self-lensing nor occultation effects. In the next section, we evaluate this point by investigating the maximum deviation in the normalized source flux analytically. 

\noindent In Figures \ref{lightc} and \ref{lightd}, we display two examples of self-lensing signals that are canceled out with the occultation effects. In both cases $R_{\rm WD}\simeq 1.4 R_{\rm E}$, whereas they have different values for $\rho_{\star}$. If such stellar light curves are discovered, the equation $R_{\rm{WD}}\simeq 1.4 R_{\rm E}$ will offer a relation between the WD mass and the orbital period (we assume the source star's parameters including its mass are determined from other observations, e.g., spectroscopy). If the eclipsing signal due to transiting the compact object by the source star happens, it reveals the orbital period. Accordingly, one can indicate the WD's mass. 

In the two last panels, \ref{lighte} and \ref{lightf}, we represent two more self-lensing events affected by finite-lens effects. In panel \ref{lighte}, the self-lensing signal is higher than the occultation effect, and in the next panel, the occultation is deeper than the self-lensing signal.

Stellar light curves shown in Figure \ref{Light2} have similar shapes. Several observing parameters can be extracted from them which are functions of the physical parameters of the source stars and compact objects. For each light curve, these observing parameters are (i) the peak value, (ii) width, (iii) the deviation from a top-hat model, (iv) the time of the maximum time-derivative of the magnification factor \citep[see, e.g.,  ][]{2022ApJJohnson,2023MNRASSajadian}. 

\noindent In Subsection \ref{sec4_1}, we offer an analytical relation for the peaks of overlapping self-lensing/occultation light curves and study the dependence of their values on relevant physical parameters. In Subsection \ref{sec4_2}, we evaluate the errors by using the offered analytical and approximate relations to estimate the peak values instead of their real values derived by the IRS method. We evaluate the widths of self-lensing/occultation light curves as explained in Subsection \ref{sec4_3}. %We do not evaluate the two last observing parameters, because they need to simulate full light curves using the IRS method which is too time-consuming. 

\subsection{Peaks of Light Curves: Estimation}\label{sec4_1}
In a self-lensing event and when the lens is over the source's disk and as far as $R_{\rm E}\ll R_{\star, \rm p}$, the image of the source star is a thick ring where its inner and outer edges are $R_{1}\simeq R_{\rm E}^{2}/R_{\star, \rm p}$ and $R_{2}\simeq R_{\star, \rm p}+ R_{\rm E}^{2}/R_{\star, \rm p}$. The inner radius is much smaller than the Einstein radius, and in most WDMS binary systems it is smaller than the WDs' radii. To show this point, the maps of $\log_{10}[R_{\rm{WD}}/R_{1}]$ over the 2D space $M_{\rm{WD}}(M_{\sun})-\log_{10}[M_{\star}(M_{\sun})]$ are shown in three top panels of Figure \ref{mapp} by considering three orbital periods. Accordingly, only for long orbital periods $T\gtrsim 150$ days and when WDs are massive, $R_{\rm{WD}}$ is less than $R_{1}$, and the occultation does not change the light curves' peak. In WDMS binary systems with low-mass WDs ($M_{\rm WD}\lesssim 0.2 M_{\sun}$), radii of WDs are significantly higher than $R_{1}$ (hundred times or even more depending on the orbital period).  

For an overlapping self-lensing/occultation signal, the peak in the stellar normalized flux can be given by:  
\begin{eqnarray}
A_{\rm{m}}=A(\rho_{\star}, \Gamma)-\mathcal{O}(R_{\rm WD},~R_{1},~R_{\star, \rm{p}}),
\label{amax}
\end{eqnarray}
where, $A(\rho_{\star}, \Gamma)$ is the maximum magnification factor. For a uniform source star, it can be estimated by $1+2~\rho_{\star}^{-2}$. Also, $\mathcal{O}=\Theta\big(R_{\rm WD}^{2}-R_{1}^{2}\big)\big/R_{\star, \rm p}^{2}$ displays the occultation amount for a uniform source star and is the ratio of the images' area which is blocked by the lens to the source area. Here, $\Theta$ is a step function which is zero if $R_{\rm WD}\leq R_{1}$. If $R_{\rm{WD}}>R_{1}$ the value of the step function is $R_{\rm WD}^{2}-R_{1}^{2}$. Three middle panels of Figure \ref{mapp} represent the maps of $\mathcal{O}(\times 100)$ over the 2D space $M_{\rm{WD}}(M_{\sun})-\log_{10}[M_{\star}(M_{\sun})]$ and for three values of the orbital period. If $R_{\star, \rm p}\gg R_{\rm E}$, one can find this analytical relation for $A_{\rm{m}}$ as follows:   
\begin{eqnarray}
&A_{\rm{m}}&=1+\Delta A_{\rm{m}}\simeq\nonumber\\
&1&+\frac{2 R_{\rm E}^{2}-R_{\rm WD}^{2}}{R_{\star, \rm{p}}^{2} }+\frac{R_{\rm E}^{4}}{R_{\star, \rm p}^{4}}-2\frac{R_{\rm E}^6}{R_{\star, \rm p}^{6}}. 
\label{amax2}
\end{eqnarray}
The higher terms, as far as $R_{\star, \rm p}\gg R_{\rm E}$, are much smaller and ignorable. Similar approximate relations for the maximum deviation while self-lensing were introduced before \citep{2003ApJEric,2016ApJHan}. Here, we calculate its amount for different WDMS systems and discuss their properties. In real observations, the orbital period is uniquely determined because self-lensing and eclipsing signals are repeated with the period exactly equal to the orbital period. Hence, the masses of two companions (star and WD) are the free parameters that should be derived from the observed peak amount. We assume their radii are estimated using the known mass-radius relations (e.g., Equation \ref{wdrm}). We calculate $\Delta A_{\rm{m}}\times 100$ numerically and display its maps over that 2D space in three bottom panels of Figure \ref{mapp}. Some points from these maps are listed in the following.  
\begin{figure}
\centering
\includegraphics[width=0.49\textwidth]{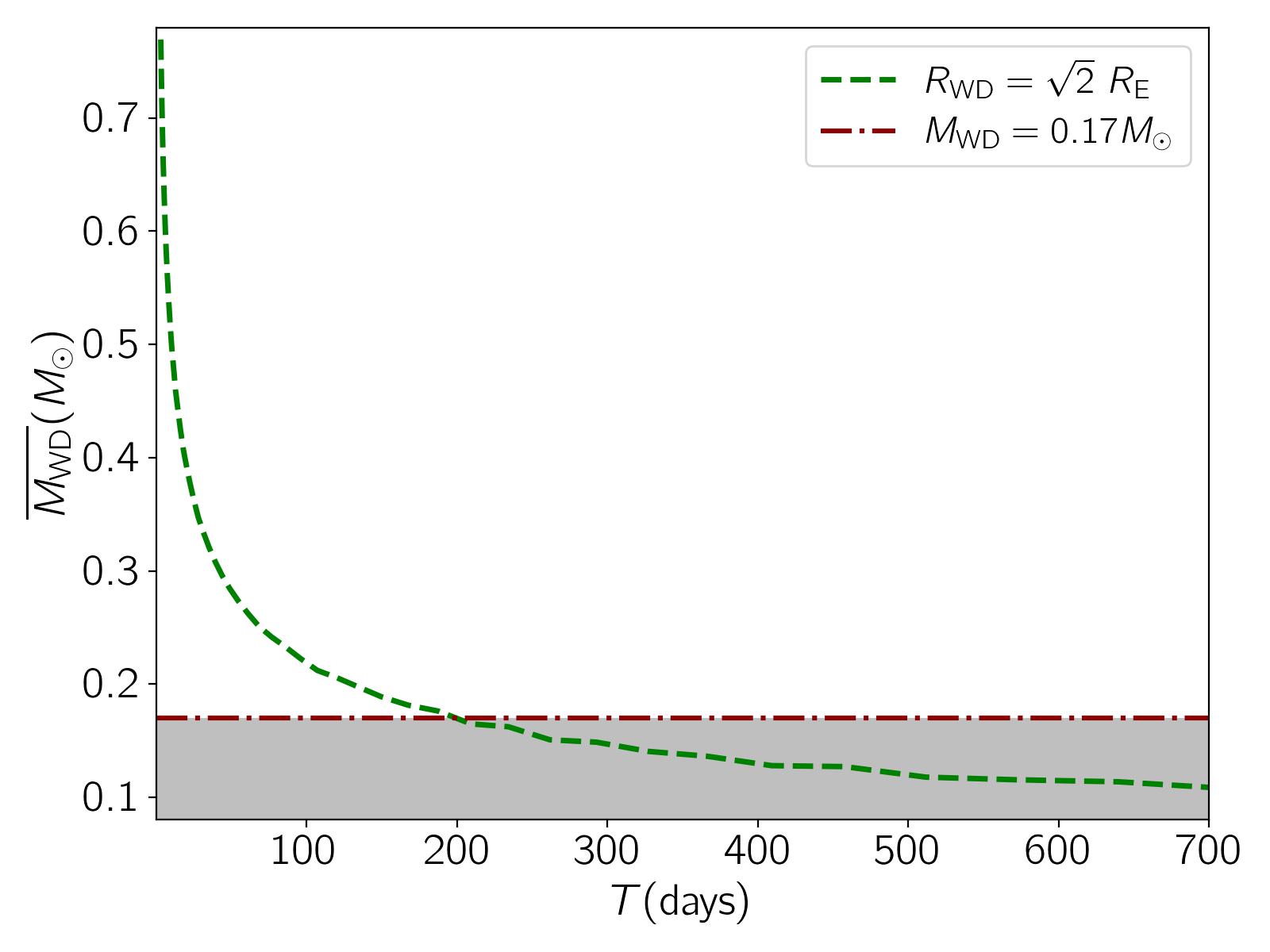}
\caption{The average mass of WDs in WDMS binary systems for which $R_{\rm{WD}}=\sqrt{2}~R_{\rm E}$ (where the self-lensing and occultation effects cancel each other out) versus the orbital period is plotted with a green dashed curve. The low limit on the WDs' mass specified based on real observations (i.e., $0.17M_{\odot}$ \citep[see, e.g.,][]{2007ApJKilic,2018AACalcaferro,2020ApJBrown}) is represented with a red dot-dashed line.}\label{cancel}
\end{figure}

\begin{itemize}[leftmargin=2.0mm]
\item Finite-lens effects are considerable when (i) the orbital period is short (some ten days) and (ii) both WD and its companion star are low-mass objects. We note that $R_{\rm{WD}}/R_{1}$ (or $R_{\rm{WD}}/R_{\rm E}$ as shown in the left panel of Figure \ref{RcNorm}) increases by $M_{\star}$. Nevertheless, the occultation effect on the stellar light curve, $\mathcal{O}$, decreases by $M_{\star}$, as it reversely depends on $R_{\star, \rm p}^{2}$.

\item The most massive WDs ($M_{\rm{WD}}\simeq 1.3M_{\sun}$) produce large self-lensing signals which are dominant to their finite-lens effects even by considering wide ranges for the orbital period, and the mass of companion stars. However, $\Delta A_{\rm{m}}$ increases by enhancing the orbital period.  

\item Also, for a given WD and its companion star $\Delta A_{\rm{m}}$ increases $3$-$5$ times if their orbital period increases by one order of magnitude.     

\item The contour line $\Delta A_{\rm{m}}=0$ (which are highlighted with magenta colour on three bottom maps) shows the WDMS binary systems in their stellar light curves self-lensing and finite-lens effects cancel each other out. This contour line happens when $R_{\rm{WD}}\simeq \sqrt{2}R_{\rm E}$ where the main term in $\Delta A_{\rm{m}}$, as given by Equation \ref{amax2}, vanishes. These contour lines for each orbital period happen almost at a fixed $M_{\rm{WD}}$ and very weakly depend on $M_{\star}$. For instance, when $T=10,~100,~300$ days, $\Delta A_{\rm{m}}=0$ happen when $M_{\rm{WD}}(M_{\sun})\simeq 0.49,~0.22,~0.14$. For a wide range of the orbital period, we numerically calculate $\overline{M_{\rm{WD}}}$ for which $\Delta A_{\rm{m}}=0$ as shown in Figure \ref{cancel} with a green dashed line. The lowest-mass WDs have been detected up to now have the mass $0.17 M_{\odot}$, \citep[see, e.g., ][]{2007ApJKilic,2018AACalcaferro,2020ApJBrown}. This low-limit on the WDs' mass is specified with a red dot-dashed line in this figure. This line intersects with the line $R_{\rm{WD}}=\sqrt{2}~R_{\rm E}$ at $T\simeq 200$ days. It means that in stellar light curves due to edge-on and wide WDMS binary systems with $T\gtrsim 200$ days, self-lensing and occultation signals do not cancel each other out completely.

\item Three bottom maps reveal the degeneracy while extracting $M_{\rm{WD}}$, and $M_{\star}$ from the maximum deviation in the normalized source flux during self-lensing/occultation $\Delta A_{\rm{m}}$. The contour curves show different values of $M_{\rm WD}$ and $M_{\star}$ that offer a given $\Delta A_{\rm{m}}$. However, the dependence (the variation rate) of $\Delta A_{\rm{m}}$ on the WD mass is stronger than that of the mass of the companion star.   

\item When the self-lensing signal dominates the finite-lens effect (i.e., $\Delta A_{\rm{m}}>0$) two parameters $M_{\rm WD}$, and $M_{\star}$ have reverse effects on $\Delta A_{\rm{m}}$ so that either increasing the first one or decreasing the second enhances $\Delta A_{\rm{m}}$.  

\item When the occultation due to the finite-lens effect dominates the self-lensing signal (i.e., $\Delta A_{\rm{m}}<0$), decreasing either $M_{\rm{WD}}$ or $M_{\star}$ increases the occultation's depth. Therefore, the masses of WD and its companion star have the same effect on the occultation signals.  

\item By considering the wide ranges for WD and companion star masses ($M_{\rm WD}\in [0.1,~1.4] M_{\sun}$, and $M_{\star}\in [0.08,~1.2] M_{\sun}$) and the orbital period ($T\in [3,~800]$ days) which are used to make maps/animations, we find the peak of overlapping self-lensing/occultation light curves mostly occurs in the range $A_{\rm m}\sim [0.975,~1.5]$. 
\end{itemize}

\subsection{Peaks of Light Curves: Error}\label{sec4_2}
\citet{2003ApJEric} first offered an analytical solution for the magnification factor of a large source star by considering a limb-darkened profile for its brightness and the occultation effect due to the finite-lens size. His analytical relation describes the peaks in self-lensing/occultation events well, as far as $\rho_{\star}\gg 1$. For that reason, in five discovered self-lensing events, modelling of light curves was done based on his formula. In most self-lensing events, $\rho_{\star}$ is too large except for WDMS binary systems with massive WDs in wide orbits (with long orbital periods). 

\begin{figure*}
\centering
\includegraphics[width=0.32\textwidth]{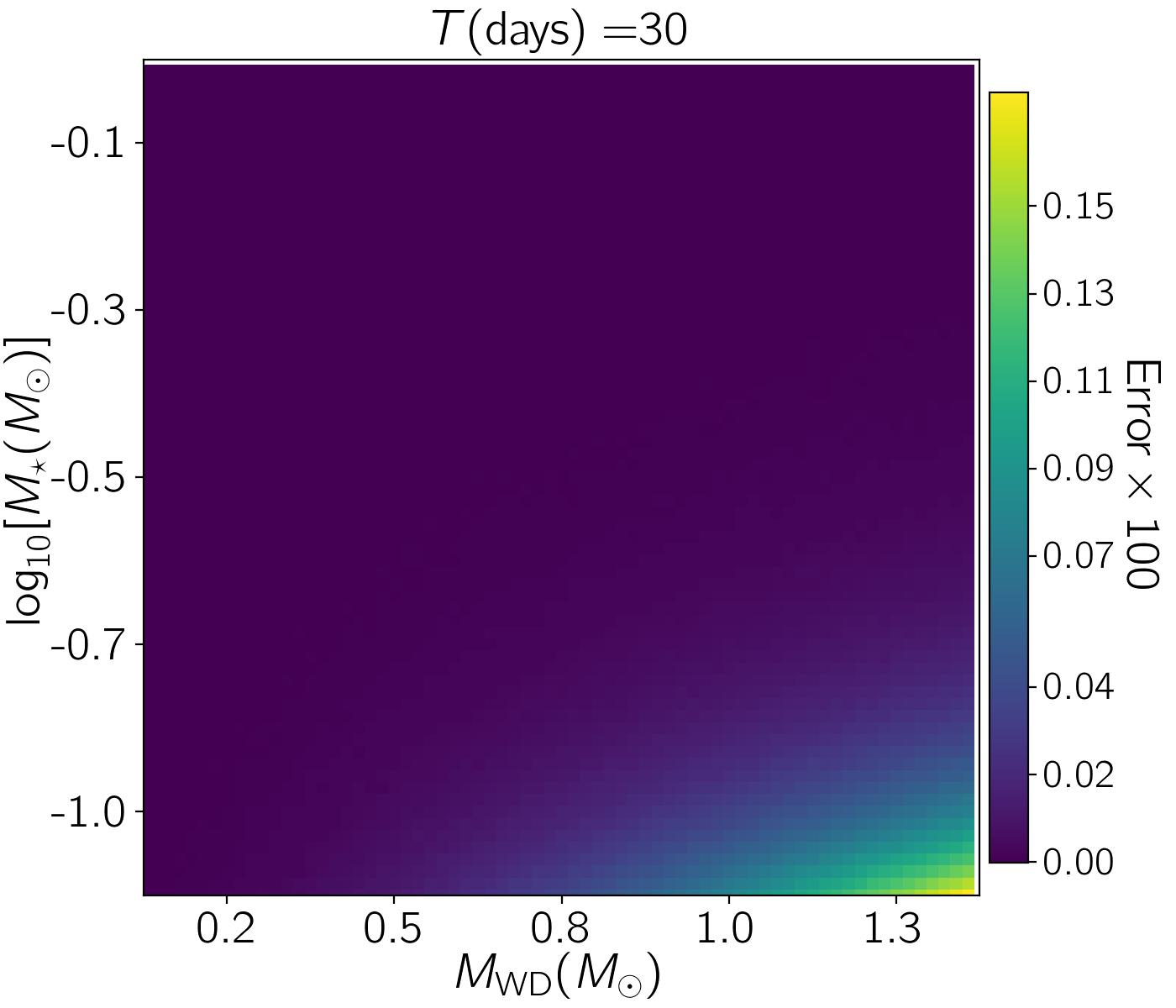}
\includegraphics[width=0.32\textwidth]{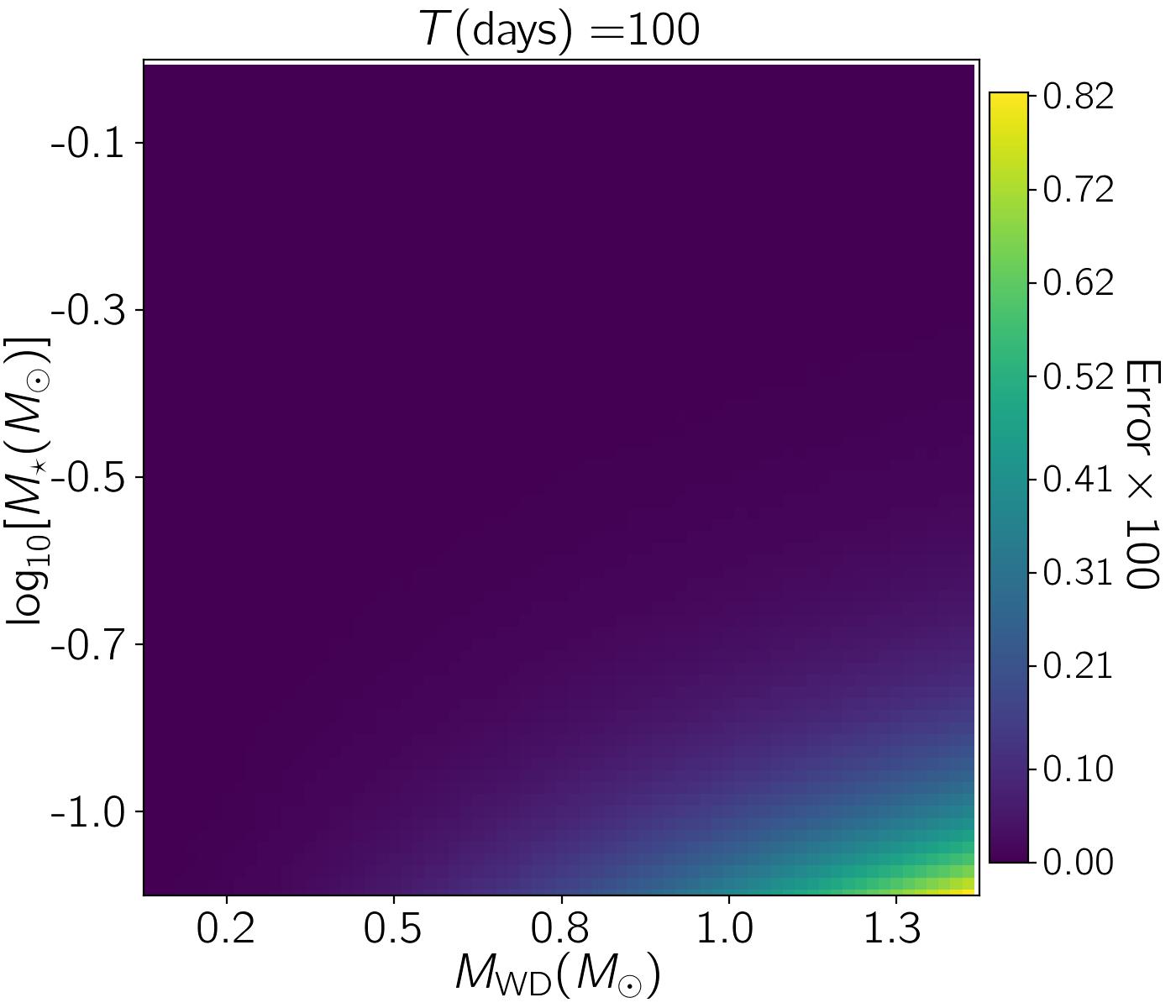}
\includegraphics[width=0.32\textwidth]{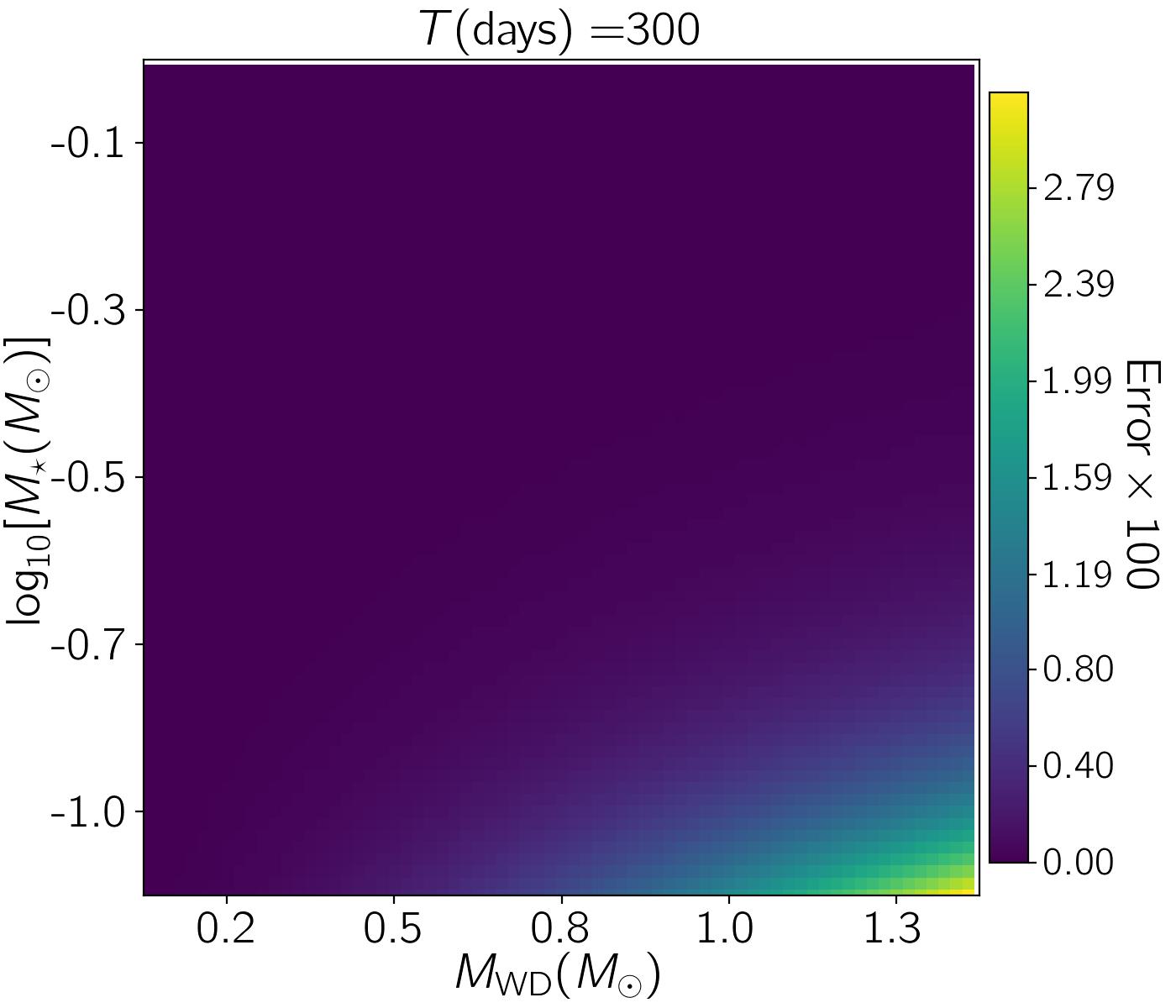}
\includegraphics[width=0.32\textwidth]{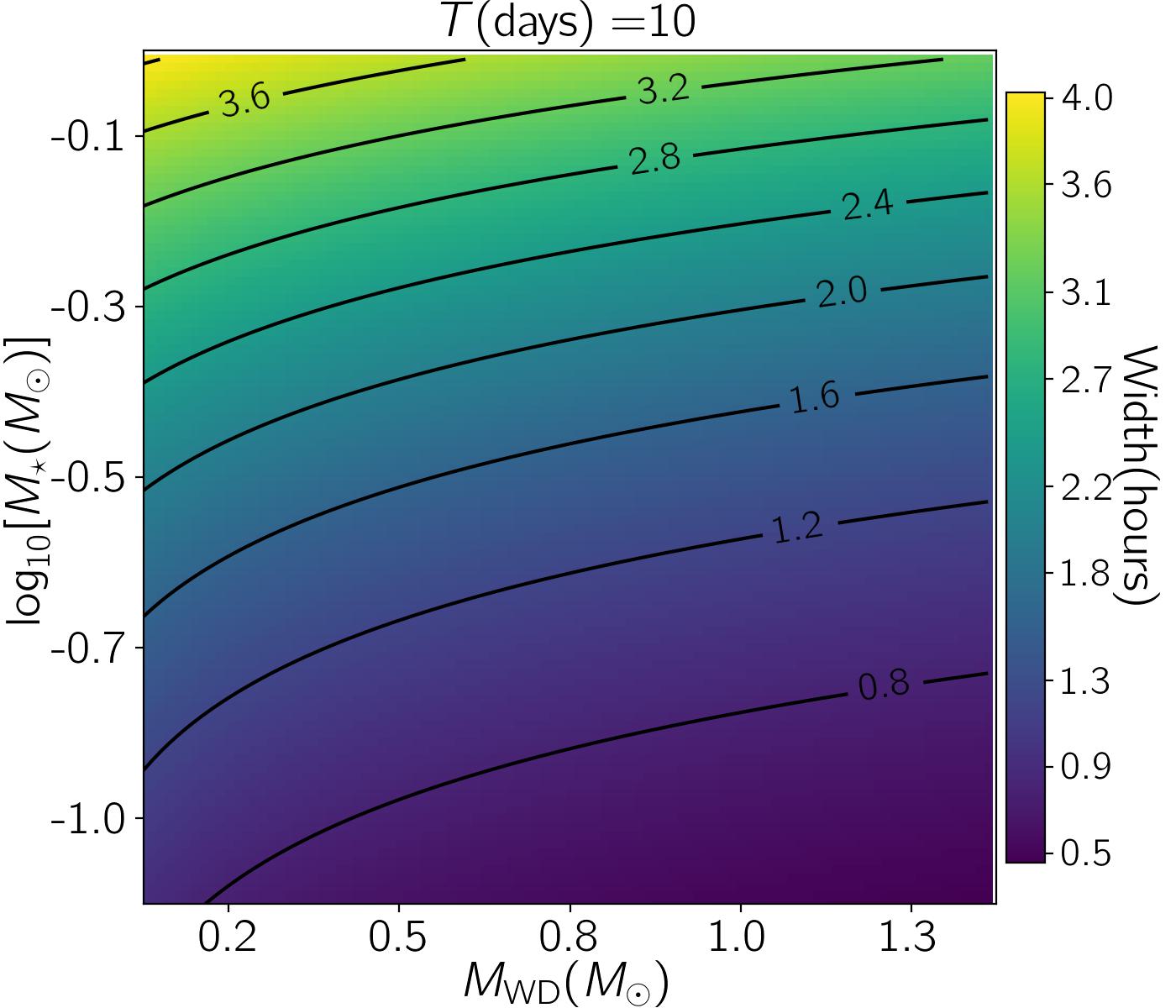}
\includegraphics[width=0.32\textwidth]{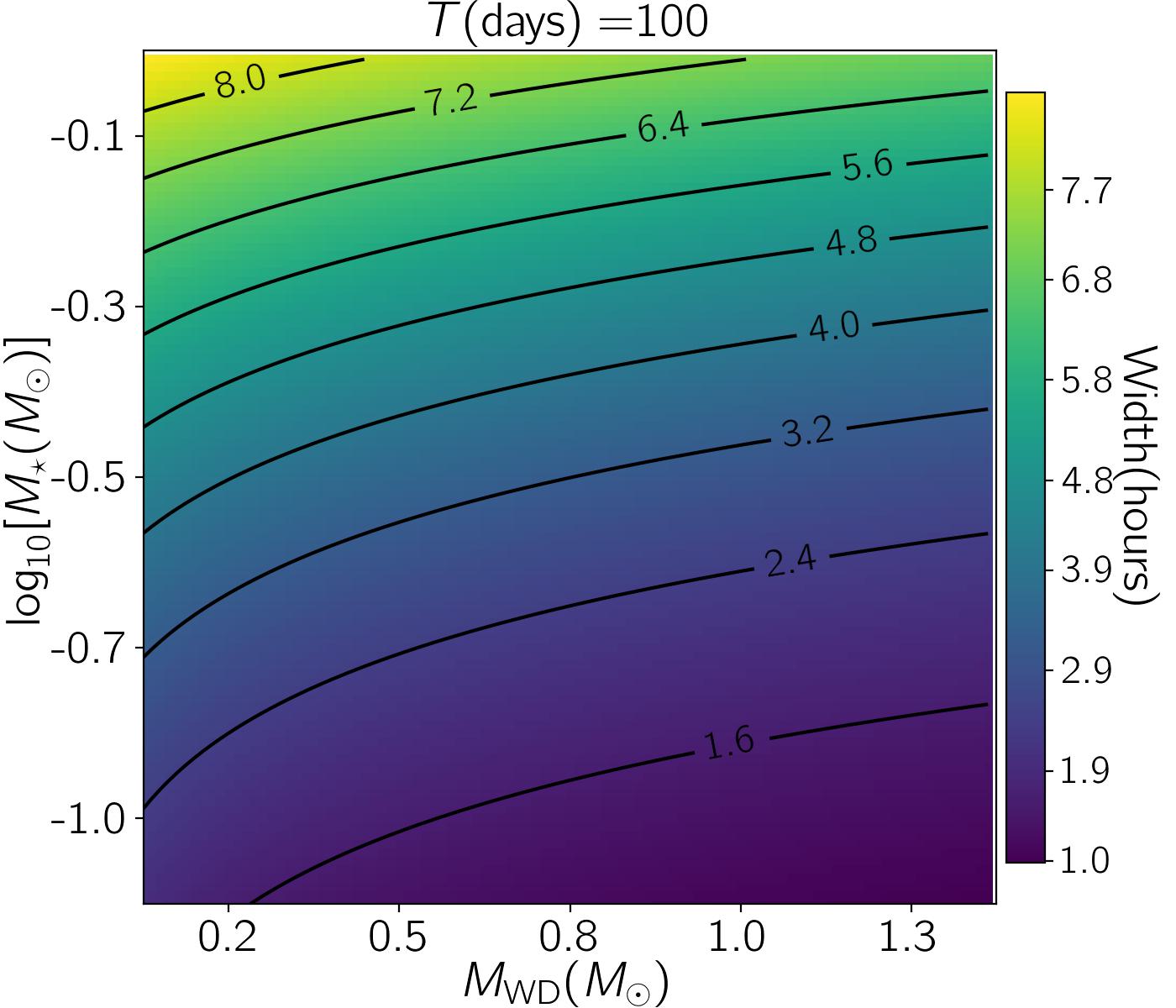}
\includegraphics[width=0.32\textwidth]{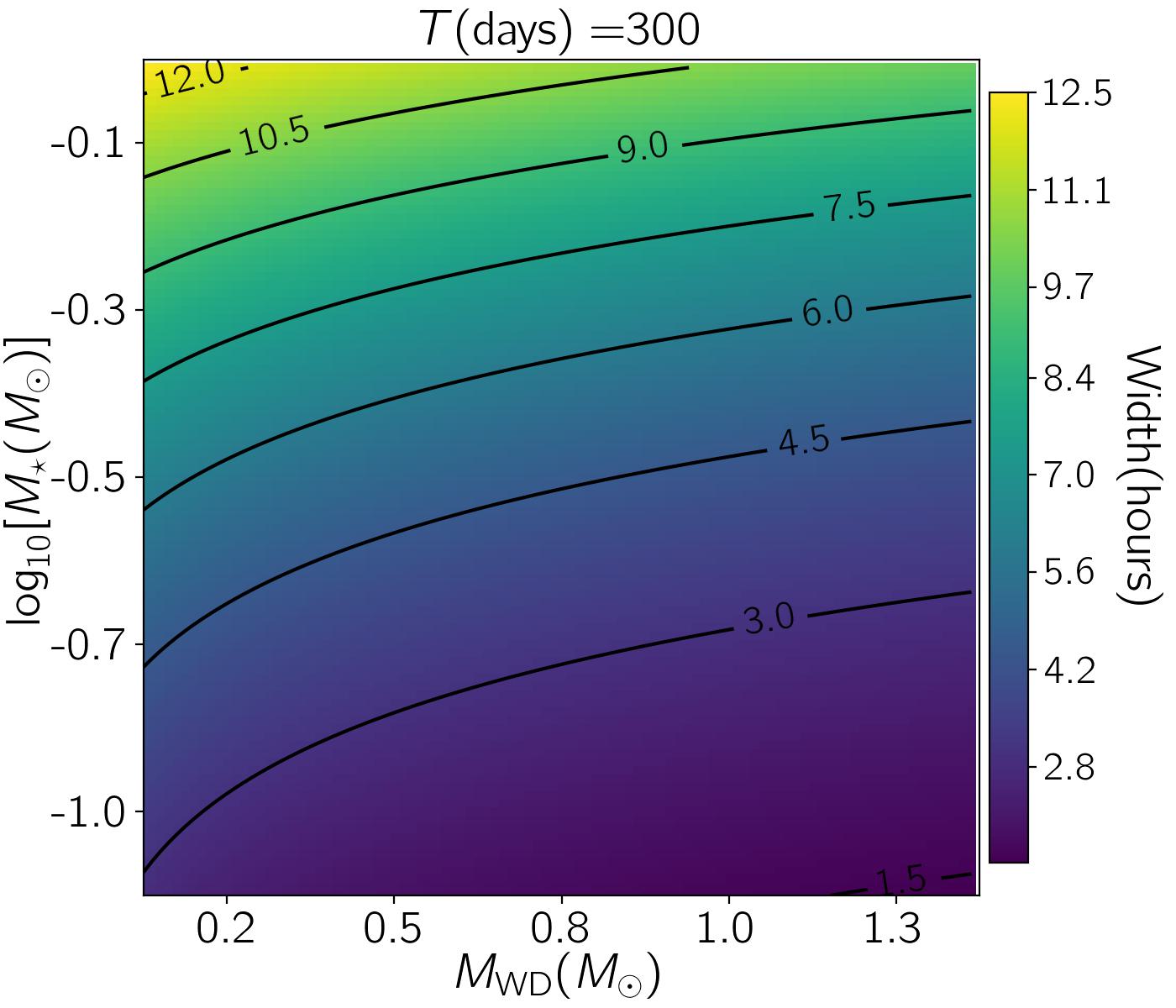}
\caption{Top panels: maps of overestimation for the maximum deviations in the normalized source flux while self-lensing ($\Delta A_{\rm m}$) owing to using Equation \ref{amax2} as compared to their real values extracted by the IRS method. Bottom panels: Maps of the width of self-lensing/occultation light curves (given by Equation \ref{wid}) from different edge-on WDMS binary systems by considering three values for the orbital period. An \href{https://iutbox.iut.ac.ir/index.php/s/bb2fy7NEjwGEk2k}{animation} from the maps of light curves' width versus the orbital period is available.}\label{error}
\end{figure*}

Here, we evaluate the errors of using analytical relations while modelling self-lensing signals. We calculate peaks of overlapping self-lensing/occultation signals in different WDMS binary systems by considering a uniform stellar brightness two times using (i) the analytical relation given by Equation \ref{amax} (an approximate value) and (i) the IRS method (an accurate value $\Delta A_{\rm{m},\rm{IRS}}$). The maps of their differences (Error=$\Delta A_{\rm{m}}- \Delta A_{\rm{m},\rm{IRS}}$) for three values of the orbital period are displayed in top panels of Figure \ref{error}. According to these maps, for analyzing the self-lensing/occultation signals when the companion stars are less massive and WDs are massive, using the approximate formula for the magnification (given by Equation \ref{amax}) could overestimate its peak up to $0.0018,~0.0082,~0.0319$ when the orbital period is $T=30,~100,~300$ days, respectively. 

\noindent We note that these errors happen for WDMS binary systems including ultra massive WDs ($M_{\rm{WD}}\gtrsim 1.2 M_{\odot}$) and low-mass stars ($M_{\star}\lesssim 0.15 M_{\odot}$). Sub-stellar companions even giant planets orbiting/transiting WDs are common and can be detected through either eclipsing or infrared/near-infrared flux excess \citep[see, e.g., ][]{2005ApJSFarihi,2020NaturVanderburg, 2022MNRASKosakowski,2023MNRASJimenez,2024MNRASFerreira}. Nevertheless, the mass distribution of WDs maximizes at $\sim 0.6 M_{\odot}$ with a low number of detected ultra massive WDs up to now \citep[see, e.g.,][]{2008AJDe}. So the discussed errors in modeling self-lensing signals occur rarely, because ultra massive WDs are not common. The only detected sub-stellar object around an ultra massive WD was reported recently by \citet{2024Chengarxiv}. They found one giant planetary companion for one of 3268 massive white dwarfs reported in the Gaia Data Release 3 (DR3, \citet{2023AAGDR3}).
    
%{According to the last map in Figure \ref{mapp}, $\Delta A_{\rm m}$ due to wide WDMS binary systems with massive WDs ($T\sim 300$ days and $M_{\rm{WD}}\gtrsim M_{\odot}$) can reach to $1.15-1.25$. Hence, for such events the error of using analytical relations would be $2.2-2.6\%$.} 

Also, according to top panels of Figure \ref{error}, this error will be higher for wider binary systems. Three discovered self-lensing events (whose Kepler's IDs are KIC 03835482, KIC 06233093, KIC 12254688) by \citet{2018AJKawahara} have long orbital periods $T\simeq 683,~728,~419$ days, but their WDs and companion stars had masses $M_{\rm WD}\simeq 0.52,~0.53,~0.5M_{\odot}$ and  $M_{\star}\simeq 1.3,~1.1,~1.4 M_{\odot}$, respectively. Since, their WDs' mass is half of solar mass no significant errors happened while modeling these events due to using analytical relations. Nevertheless, for these three targets we evaluated $\Delta A_{\rm{m},\rm{IRS}}$ which resulted in $\rm{Error}\simeq 4.0e-4, 5e-4, 2e-4$, respectively.

\subsection{Widths of Light Curves}\label{sec4_3}
The overlapping self-lensing/occultation light curves that are shown in Figure \ref{Light2} are similar to top-hat models, and their deviations from top-hat models are not large. Hence, we define their width as the time to cross the source disk by the lens object (which is somewhat different from the Full Width at Half Maximum), as given by: 
\begin{eqnarray}
\rm{width}= 2 t_{\star},~~~t_{\star}=\frac{T}{2 \pi}\Big|\frac{\pi}{2}-\arccos \frac{R_{\star, \rm{p}}}{a}\Big|.
\label{wid}
\end{eqnarray}
Here, we ignore the limb-darkening effect for the stellar brightness profile, while considering circular and completely edge-on orbits. As far as $R_{\star, \rm p}\ll a$ for detached WDMS binary systems, we can expand the term $\arccos$ which results: 
\begin{eqnarray}
\rm{width}\simeq \frac{T}{\pi}\Big[\frac{R_{\star, \rm p}}{a}+\frac{1}{6}\big(\frac{R_{\star, \rm p}}{a}\big)^{3}+...\Big],
\end{eqnarray}
Accordingly, the width of self-lensing/occultation signals is proportional to $\propto T^{1/3} R_{\star, \rm p} (M_{\star}+M_{\rm{WD}})^{-1/3}$. We estimate the width of different stellar self-lensing/occultation light curves due to WDMS binary systems using Equation \ref{wid}. The maps of these widths for three orbital periods and over 2D space $M_{\rm{WD}}(M_{\sun})-\log_{10}[M_{\star}(M_{\sun})]$ are represented in three bottom panels of Figure \ref{error}. As mentioned, the light curves' width increases by $T^{1/3}$. For a given orbital period, smaller source stars make deeper self-lensing/occultation signals (see Equation \ref{amax2}), but with smaller widths. Similarly, more massive WDs make deeper signals, but with smaller widths compared to less massive WDs.    

\begin{figure}
\centering
\includegraphics[width=0.49\textwidth]{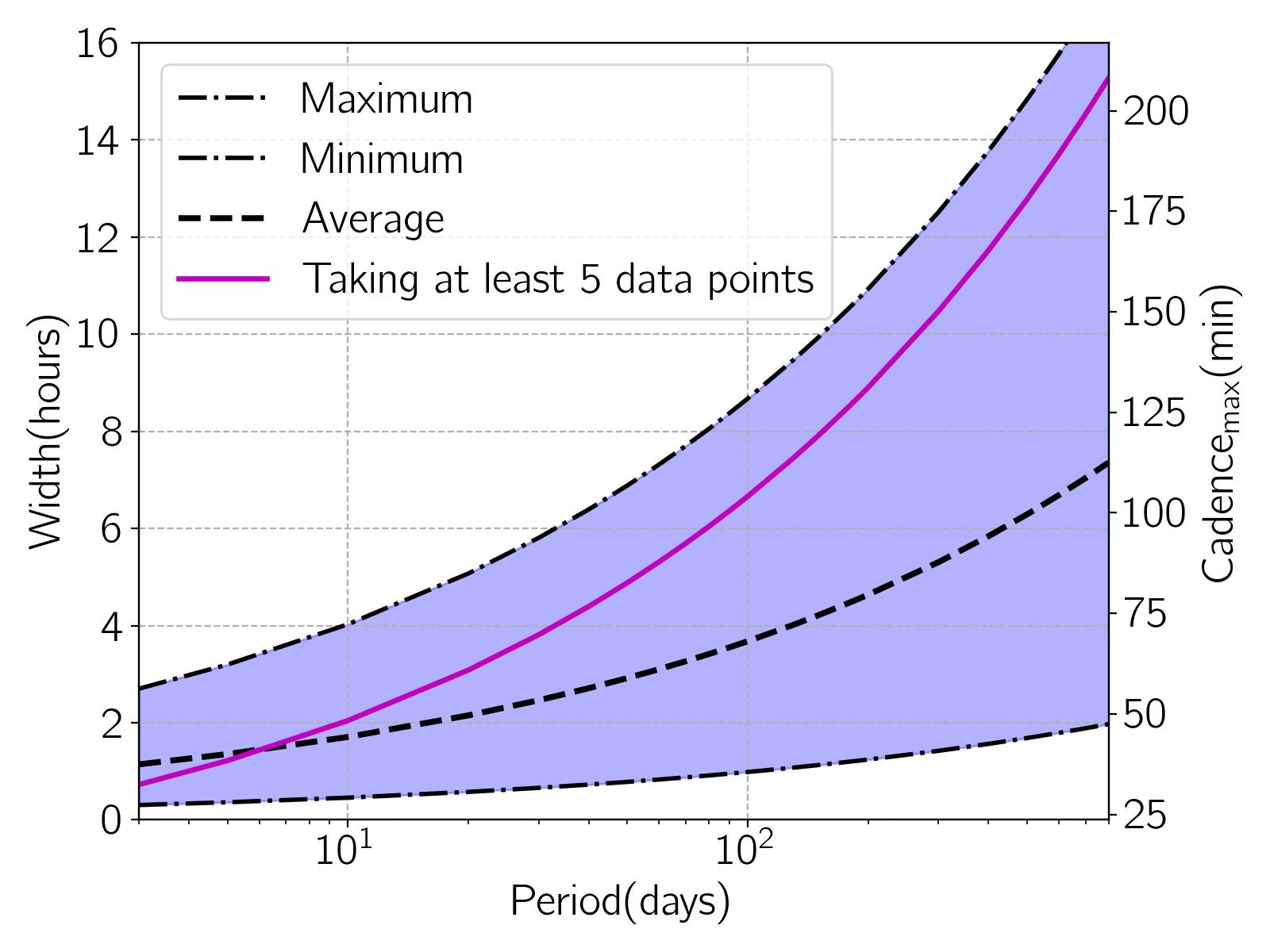}
\caption{The ranges and the average of possible widths (specified on the left vertical axis) due to different self-lensing/occultation signals from different WDMS binary systems are shown with dot-dashed and dashed black curves versus the orbital period, respectively. We assume at least 5 data points should cover signals so that they can be realized. In that case, the worst cadence versus the orbital period for realizing these self-lensing/occultation signals is $\rm{Cadence}_{\rm{max}}(\rm{minute})=\rm{Maximum}\times12$ and plotted with a magenta solid line as specified on the right vertical axis.}\label{updown}
\end{figure}

To have a sense of widths of possible self-lensing/occultation light curves for different orbital periods, in Figure \ref{updown} we show the range of all possible width values (dot-dashed black curves) and their average (dashed black curve) versus the orbital period as specified on the left vertical axis. If we assume for discerning any self-lensing/occultation signal at least 5 data points should be recorded, the worst cadence to capture the signals with a given orbital period is $\rm{Cadence}_{\rm{max}}(\rm{minute})=\rm{Maximum}(T)\times 12$, which is shown in this figure with a magenta solid line and specified on the right vertical axis. Accordingly, any survey observations with a cadence worse than $110$, $50$, and $30$ minutes could not find any self-lensing/occultation signals due to WDMS binary systems with orbital periods less than $\sim 100$, $10$, and $3$ days, respectively. To evaluate the light curves' width in Equation \ref{wid}, we assumed (i) the inclination angle is zero, (ii) the stellar brightness is uniform without any limb-darkening effect, and (iii) the stellar orbit around the compact object is circular with a constant speed. Two first simplifications lead to an overestimation for the light curves' width. Hence with the determined $\rm{Cadence}_{\rm{max}}$ we may miss some short self-lensing/eclipsing signals due to limb-darked source star and higher inclination angle.

\section{Conclusions}\label{sec5}
In this work, we studied the finite-lens effect in the self-lensing signals due to edge-on and detached WDMS binary systems analytically (based on a close approximation to reality) and numerically (using the inverse ray shooting (IRS) method). Finite-lens effect refers to the obscuration of some part of the images' area by the lens's disk. In WDMS binary systems with orbital radii around one astronomical unit, their Einstein radii are $\sim0.02$-$0.05R_{\sun}$ which is in the same order of magnitude with the radius of a common WD. Hence, during self-lensing signals due to WDMS binary systems, the compact object can block some part of the images' disk which are inside the Einstein ring. 

To find in what WDMS systems this occultation effect is considerable, we first calculated $R_{\rm WD}/R_{\rm E}$ for different masses of WDs and source stars versus the orbital period. We concluded that for most of WDs and source stars with $M_{\rm WD}\gtrsim 0.2 M_{\sun}$, and $M_{\star}\in [0.08,~1.2] M_{\sun}$ if their orbital periods are $T\gtrsim 300$ days, the magnification due to self-lensing dominates to the occultation owing to the finite-lens effect. Most massive WDs with $M_{\rm WD}\gtrsim 1.3M_{\odot}$ with very small radii, even by considering a wide range for the orbital period $T \in [5,~1000]$ days and different values for $M_{\star}$, we do not expect significant finite-lens effect, since for them $R_{\rm{WD}}\lesssim0.6R_{\rm E}$. 

Generally, self-lensing signals dominate finite-lens effects when $R_{\rm E}$ is large. The highest self-lensing signal happens for most massive WDs ($M_{\rm{WD}}\sim 1.3 M_{\odot}$) that have dwarf companions in wide orbits. Inversely, the occultation dominates the self-lensing signal when $R_{\rm E}$ is small, e.g., for WDMS systems with low-mass WDs ($M_{\rm WD}\sim 0.2 M_{\sun}$) and dwarf companion stars in close orbits with short orbital periods $T\lesssim 50$ days. We noted that both effects (the magnification and occultation) enhance when the source radius reduces. Also, for long orbital periods $T\gtrsim150$ days and when WDs are massive, there is no occultation effect. 

We also confirmed these results by evaluating the occultation amounts $\mathcal{O}$ and the maximum deviations in the stellar normalized flux (due to overlapping self-lensing and occultation effects) for different WDMS binary systems, $\Delta A_{\rm{m}}$  based on the analytical and approximate relations. Considering the wide ranges for WD and companion star masses and the orbital period which are applied in the making maps/animations, the peaks of self-lensing/occultation light curves mostly occur in the range $\Delta A_{\rm m}\in [0.975,~1.5]$. 

Using the IRS method, we made stellar light curves for detached and edge-on WDMS binary systems and when the source star passes behind the compact object. In these light curves, self-lensing and occultation due to finite-lens size canceled each other out when $R_{\rm WD}\simeq \sqrt{2}R_{\rm E}$, regardless of the source radius. This condition happens for a certain $M_{\rm{WD}}$ for any given value of the orbital period. For instance self-lensing and occultation cancel each other out when $M_{\rm WD}(M_{\sun})\sim 0.49,~0.22,~0.14$ when the orbital period is $T=10,~100,~300$ days, respectively. Considering the fact that the lowest-mass WDs detected up to now have the mass $0.17M_{\odot}$ \citep[][]{2007ApJKilic,2018AACalcaferro,2020ApJBrown}, for the WDMS binary systems with $T\gtrsim 200$ days the cancellation of self-lensing and occultation signals never happens.

Because all five self-lensing events discovered up to now \citep{KruseAgol2014,2018AJKawahara,2019ApJLMasuda,2024Yamaguchi} were modeled using the known analytical and approximate relation (given in Equation \ref{amax2}), we evaluated the values of overestimation in $\Delta A_{\rm{m}}$ with respect to the real amounts extracted by the IRS method because of using the analytical relation. The maps of these deviations were shown in three top panels of Figure \ref{error} for three given values of the orbital period. According to these maps, for analyzing the self-lensing signals using the approximate formula for the magnification could overestimate its peak up to $0.0018,~0.082,~0.0319$ when the orbital period is $T=30,~100,~300$ days, respectively. These maximum overestimations occur for WDMS systems including ultra massive WDs ($M_{\rm{WD}}\gtrsim 1.2 M_{\odot}$) and sub-stellar companions ($M_{\star}\lesssim 0.15 M_{\odot}$). None of those five discovered self-lensing events were such systems. Generally, the number of detected ultra massive WDs is rare.

Finally, we studied the dependence of the stellar light curves' width to the physical parameters which is $\rm{width}\propto R_{\star, \rm p}~T^{1/3} (M_{\star}+M_{\rm WD})^{-1/3}$. Accordingly, smaller source stars make deeper signals, but with shorter durations in comparison with larger ones. We considered wide ranges for masses of WDs and main-sequence source stars in completely edge-on orbits and with a uniform brightness for stellar disks, and estimated widths of their light curves. Based on these assumptions, we concluded any survey observations with a cadence worse than $110$, $50$, $30$ minutes could not find any self-lensing/occultation signals due to WDMS binary systems with the orbital period less than $100$, $10$, $3$ days, respectively. However, even with better cadences some signals due to more-inclined systems and limb-darkened source stars may be missed.

\small{The authors thank the anonymous referee for his/her careful and useful comments.}

\small {All simulations that have been done for this paper are available at:  \url{https://github.com/SSajadian54/FiniteLensEffect}. Also, the codes, animations, figures and several examples of generated light curves can be found in the Zenodo repository\citep{sajadian2024b}.}\\

\bibliographystyle{aasjournal}
\bibliography{ref}{}

\begin{thebibliography}{}
\expandafter\ifx\csname natexlab\endcsname\relax\def\natexlab#1{#1}\fi
\providecommand{\url}[1]{\href{#1}{#1}}
\providecommand{\dodoi}[1]{doi:~\href{http://doi.org/#1}{\nolinkurl{#1}}}
\providecommand{\doeprint}[1]{\href{http://ascl.net/#1}{\nolinkurl{http://ascl.net/#1}}}
\providecommand{\doarXiv}[1]{\href{https://arxiv.org/abs/#1}{\nolinkurl{https://arxiv.org/abs/#1}}}

\bibitem[{{Agol}(2002)}]{2002ApJAgol}
{Agol}, E. 2002, \apj, 579, 430, \dodoi{10.1086/342880}

\bibitem[{{Agol}(2003)}]{2003ApJEric}
---. 2003, \apj, 594, 449, \dodoi{10.1086/376833}

\bibitem[{{Ambrosino}(2020)}]{2020Ambrosino}
{Ambrosino}, F. 2020, arXiv e-prints, arXiv:2012.01242,
  \dodoi{10.48550/arXiv.2012.01242}

\bibitem[{{Brown} {et~al.}(2020){Brown}, {Kilic}, {Kosakowski}, {Andrews},
  {Heinke}, {Ag{\"u}eros}, {Camilo}, {Gianninas}, {Hermes}, \&
  {Kenyon}}]{2020ApJBrown}
{Brown}, W.~R., {Kilic}, M., {Kosakowski}, A., {et~al.} 2020, \apj, 889, 49,
  \dodoi{10.3847/1538-4357/ab63cd}

\bibitem[{{Calcaferro} {et~al.}(2018){Calcaferro}, {Althaus}, \&
  {C{\'o}rsico}}]{2018AACalcaferro}
{Calcaferro}, L.~M., {Althaus}, L.~G., \& {C{\'o}rsico}, A.~H. 2018, \aap, 614,
  A49, \dodoi{10.1051/0004-6361/201732551}

\bibitem[{{Cheng} {et~al.}(2024){Cheng}, {Schlaufman}, \&
  {Caiazzo}}]{2024Chengarxiv}
{Cheng}, S., {Schlaufman}, K.~C., \& {Caiazzo}, I. 2024, arXiv e-prints,
  arXiv:2408.03985, \dodoi{10.48550/arXiv.2408.03985}

\bibitem[{Clark(1972)}]{clark1972uniform}
Clark, E.~E. 1972, Monthly Notices of the Royal Astronomical Society, 158, 233

\bibitem[{{De Gennaro} {et~al.}(2008){De Gennaro}, {von Hippel}, {Winget},
  {Kepler}, {Nitta}, {Koester}, \& {Althaus}}]{2008AJDe}
{De Gennaro}, S., {von Hippel}, T., {Winget}, D.~E., {et~al.} 2008, \aj, 135,
  1, \dodoi{10.1088/0004-6256/135/1/1}

\bibitem[{{Dominik}(1998)}]{1998Dominik}
{Dominik}, M. 1998, \aap, 329, 361, \dodoi{10.48550/arXiv.astro-ph/9702039}

\bibitem[{{Einstein}(1936)}]{Einstein1936}
{Einstein}, A. 1936, Science, 84, 506, \dodoi{10.1126/science.84.2188.506}

\bibitem[{{Farihi} {et~al.}(2005){Farihi}, {Becklin}, \&
  {Zuckerman}}]{2005ApJSFarihi}
{Farihi}, J., {Becklin}, E.~E., \& {Zuckerman}, B. 2005, \apjs, 161, 394,
  \dodoi{10.1086/444362}

\bibitem[{{Ferreira} {et~al.}(2024){Ferreira}, {Saito}, {Minniti},
  {Mej{\'\i}as}, {Caceres}, {Alonso-Garc{\'\i}a}, {Beam{\'\i}n}, {Smith},
  {Gomez}, {Lucas}, \& {Ivanov}}]{2024MNRASFerreira}
{Ferreira}, T., {Saito}, R.~K., {Minniti}, D., {et~al.} 2024, \mnras, 527,
  10737, \dodoi{10.1093/mnras/stad3911}

\bibitem[{{Gaia Collaboration} {et~al.}(2023){Gaia Collaboration}, {Vallenari},
  {Brown}, {Prusti}, {de Bruijne}, {Arenou}, {Babusiaux}, {Biermann},
  {Creevey}, {Ducourant}, {Evans}, {Eyer}, {Guerra}, {Hutton}, {Jordi},
  {Klioner}, {Lammers}, {Lindegren}, {Luri}, {Mignard}, {Panem}, {Pourbaix},
  {Randich}, {Sartoretti}, {Soubiran}, {Tanga}, {Walton}, {Bailer-Jones},
  {Bastian}, {Drimmel}, {Jansen}, {Katz}, {Lattanzi}, {van Leeuwen}, {Bakker},
  {Cacciari}, {Casta{\~n}eda}, {De Angeli}, {Fabricius}, {Fouesneau},
  {Fr{\'e}mat}, {Galluccio}, {Guerrier}, {Heiter}, {Masana}, {Messineo},
  {Mowlavi}, {Nicolas}, {Nienartowicz}, {Pailler}, {Panuzzo}, {Riclet}, {Roux},
  {Seabroke}, {Sordo}, {Th{\'e}venin}, {Gracia-Abril}, {Portell}, {Teyssier},
  {Altmann}, {Andrae}, {Audard}, {Bellas-Velidis}, {Benson}, {Berthier},
  {Blomme}, {Burgess}, {Busonero}, {Busso}, {C{\'a}novas}, {Carry}, {Cellino},
  {Cheek}, {Clementini}, {Damerdji}, {Davidson}, {de Teodoro}, {Nu{\~n}ez
  Campos}, {Delchambre}, {Dell'Oro}, {Esquej}, {Fern{\'a}ndez-Hern{\'a}ndez},
  {Fraile}, {Garabato}, {Garc{\'\i}a-Lario}, {Gosset}, {Haigron}, {Halbwachs},
  {Hambly}, {Harrison}, {Hern{\'a}ndez}, {Hestroffer}, {Hodgkin}, {Holl},
  {Jan{\ss}en}, {Jevardat de Fombelle}, {Jordan}, {Krone-Martins}, {Lanzafame},
  {L{\"o}ffler}, {Marchal}, {Marrese}, {Moitinho}, {Muinonen}, {Osborne},
  {Pancino}, {Pauwels}, {Recio-Blanco}, {Reyl{\'e}}, {Riello}, {Rimoldini},
  {Roegiers}, {Rybizki}, {Sarro}, {Siopis}, {Smith}, {Sozzetti}, {Utrilla},
  {van Leeuwen}, {Abbas}, {{\'A}brah{\'a}m}, {Abreu Aramburu}, {Aerts},
  {Aguado}, {Ajaj}, {Aldea-Montero}, {Altavilla}, {{\'A}lvarez}, {Alves},
  {Anders}, {Anderson}, {Anglada Varela}, {Antoja}, {Baines}, {Baker},
  {Balaguer-N{\'u}{\~n}ez}, {Balbinot}, {Balog}, {Barache}, {Barbato},
  {Barros}, {Barstow}, {Bartolom{\'e}}, {Bassilana}, {Bauchet}, {Becciani},
  {Bellazzini}, {Berihuete}, {Bernet}, {Bertone}, {Bianchi}, {Binnenfeld},
  {Blanco-Cuaresma}, {Blazere}, {Boch}, {Bombrun}, {Bossini}, {Bouquillon},
  {Bragaglia}, {Bramante}, {Breedt}, {Bressan}, {Brouillet}, {Brugaletta},
  {Bucciarelli}, {Burlacu}, {Butkevich}, {Buzzi}, {Caffau}, {Cancelliere},
  {Cantat-Gaudin}, {Carballo}, {Carlucci}, {Carnerero}, {Carrasco},
  {Casamiquela}, {Castellani}, {Castro-Ginard}, {Chaoul}, {Charlot}, {Chemin},
  {Chiaramida}, {Chiavassa}, {Chornay}, {Comoretto}, {Contursi}, {Cooper},
  {Cornez}, {Cowell}, {Crifo}, {Cropper}, {Crosta}, {Crowley}, {Dafonte},
  {Dapergolas}, {David}, {David}, {de Laverny}, {De Luise}, {De March}, {De
  Ridder}, {de Souza}, {de Torres}, {del Peloso}, {del Pozo}, {Delbo},
  {Delgado}, {Delisle}, {Demouchy}, {Dharmawardena}, {Di Matteo}, {Diakite},
  {Diener}, {Distefano}, {Dolding}, {Edvardsson}, {Enke}, {Fabre}, {Fabrizio},
  {Faigler}, {Fedorets}, {Fernique}, {Fienga}, {Figueras}, {Fournier},
  {Fouron}, {Fragkoudi}, {Gai}, {Garcia-Gutierrez}, {Garcia-Reinaldos},
  {Garc{\'\i}a-Torres}, {Garofalo}, {Gavel}, {Gavras}, {Gerlach}, {Geyer},
  {Giacobbe}, {Gilmore}, {Girona}, {Giuffrida}, {Gomel}, {Gomez},
  {Gonz{\'a}lez-N{\'u}{\~n}ez}, {Gonz{\'a}lez-Santamar{\'\i}a},
  {Gonz{\'a}lez-Vidal}, {Granvik}, {Guillout}, {Guiraud},
  {Guti{\'e}rrez-S{\'a}nchez}, {Guy}, {Hatzidimitriou}, {Hauser}, {Haywood},
  {Helmer}, {Helmi}, {Sarmiento}, {Hidalgo}, {Hilger}, {H{\l}adczuk}, {Hobbs},
  {Holland}, {Huckle}, {Jardine}, {Jasniewicz}, {Jean-Antoine Piccolo},
  {Jim{\'e}nez-Arranz}, {Jorissen}, {Juaristi Campillo}, {Julbe}, {Karbevska},
  {Kervella}, {Khanna}, {Kontizas}, {Kordopatis}, {Korn}, {K{\'o}sp{\'a}l},
  {Kostrzewa-Rutkowska}, {Kruszy{\'n}ska}, {Kun}, {Laizeau}, {Lambert},
  {Lanza}, {Lasne}, {Le Campion}, {Lebreton}, {Lebzelter}, {Leccia}, {Leclerc},
  {Lecoeur-Taibi}, {Liao}, {Licata}, {Lindstr{\o}m}, {Lister}, {Livanou},
  {Lobel}, {Lorca}, {Loup}, {Madrero Pardo}, {Magdaleno Romeo}, {Managau},
  {Mann}, {Manteiga}, {Marchant}, {Marconi}, {Marcos}, {Marcos Santos},
  {Mar{\'\i}n Pina}, {Marinoni}, {Marocco}, {Marshall}, {Martin Polo},
  {Mart{\'\i}n-Fleitas}, {Marton}, {Mary}, {Masip}, {Massari},
  {Mastrobuono-Battisti}, {Mazeh}, {McMillan}, {Messina}, {Michalik}, {Millar},
  {Mints}, {Molina}, {Molinaro}, {Moln{\'a}r}, {Monari}, {Mongui{\'o}},
  {Montegriffo}, {Montero}, {Mor}, {Mora}, {Morbidelli}, {Morel}, {Morris},
  {Muraveva}, {Murphy}, {Musella}, {Nagy}, {Noval}, {Oca{\~n}a}, {Ogden},
  {Ordenovic}, {Osinde}, {Pagani}, {Pagano}, {Palaversa}, {Palicio},
  {Pallas-Quintela}, {Panahi}, {Payne-Wardenaar}, {Pe{\~n}alosa Esteller},
  {Penttil{\"a}}, {Pichon}, {Piersimoni}, {Pineau}, {Plachy}, {Plum}, {Poggio},
  {Pr{\v{s}}a}, {Pulone}, {Racero}, {Ragaini}, {Rainer}, {Raiteri}, {Rambaux},
  {Ramos}, {Ramos-Lerate}, {Re Fiorentin}, {Regibo}, {Richards}, {Rios Diaz},
  {Ripepi}, {Riva}, {Rix}, {Rixon}, {Robichon}, {Robin}, {Robin}, {Roelens},
  {Rogues}, {Rohrbasser}, {Romero-G{\'o}mez}, {Rowell}, {Royer}, {Ruz Mieres},
  {Rybicki}, {Sadowski}, {S{\'a}ez N{\'u}{\~n}ez}, {Sagrist{\`a} Sell{\'e}s},
  {Sahlmann}, {Salguero}, {Samaras}, {Sanchez Gimenez}, {Sanna},
  {Santove{\~n}a}, {Sarasso}, {Schultheis}, {Sciacca}, {Segol}, {Segovia},
  {S{\'e}gransan}, {Semeux}, {Shahaf}, {Siddiqui}, {Siebert}, {Siltala},
  {Silvelo}, {Slezak}, {Slezak}, {Smart}, {Snaith}, {Solano}, {Solitro},
  {Souami}, {Souchay}, {Spagna}, {Spina}, {Spoto}, {Steele},
  {Steidelm{\"u}ller}, {Stephenson}, {S{\"u}veges}, {Surdej}, {Szabados},
  {Szegedi-Elek}, {Taris}, {Taylor}, {Teixeira}, {Tolomei}, {Tonello}, {Torra},
  {Torra}, {Torralba Elipe}, {Trabucchi}, {Tsounis}, {Turon}, {Ulla}, {Unger},
  {Vaillant}, {van Dillen}, {van Reeven}, {Vanel}, {Vecchiato}, {Viala},
  {Vicente}, {Voutsinas}, {Weiler}, {Wevers}, {Wyrzykowski}, {Yoldas}, {Yvard},
  {Zhao}, {Zorec}, {Zucker}, \& {Zwitter}}]{2023AAGDR3}
{Gaia Collaboration}, {Vallenari}, A., {Brown}, A.~G.~A., {et~al.} 2023, \aap,
  674, A1, \dodoi{10.1051/0004-6361/202243940}

\bibitem[{{Gaudi}(2012)}]{2012Gaudireview}
{Gaudi}, B.~S. 2012, \araa, 50, 411,
  \dodoi{10.1146/annurev-astro-081811-125518}

\bibitem[{{Gould}(1995)}]{1995ApJGould}
{Gould}, A. 1995, \apj, 446, 541, \dodoi{10.1086/175812}

\bibitem[{{Gould} \& {Gaucherel}(1996)}]{1996Gould}
{Gould}, A., \& {Gaucherel}, C. 1996, arXiv e-prints, astro,
  \dodoi{10.48550/arXiv.astro-ph/9606105}

\bibitem[{{Han}(2016)}]{2016ApJHan}
{Han}, C. 2016, \apj, 820, 53, \dodoi{10.3847/0004-637X/820/1/53}

\bibitem[{{Jim{\'e}nez-Esteban} {et~al.}(2023){Jim{\'e}nez-Esteban}, {Torres},
  {Rebassa-Mansergas}, {Cruz}, {Murillo-Ojeda}, {Solano}, {Rodrigo}, \&
  {Camisassa}}]{2023MNRASJimenez}
{Jim{\'e}nez-Esteban}, F.~M., {Torres}, S., {Rebassa-Mansergas}, A., {et~al.}
  2023, \mnras, 518, 5106, \dodoi{10.1093/mnras/stac3382}

\bibitem[{{Johnson} {et~al.}(2022){Johnson}, {Penny}, \&
  {Gaudi}}]{2022ApJJohnson}
{Johnson}, S.~A., {Penny}, M.~T., \& {Gaudi}, B.~S. 2022, \apj, 927, 63,
  \dodoi{10.3847/1538-4357/ac4bca}

\bibitem[{{Kawahara} {et~al.}(2018){Kawahara}, {Masuda}, {MacLeod}, {Latham},
  {Bieryla}, \& {Benomar}}]{2018AJKawahara}
{Kawahara}, H., {Masuda}, K., {MacLeod}, M., {et~al.} 2018, \aj, 155, 144,
  \dodoi{10.3847/1538-3881/aaaaaf}

\bibitem[{{Kayser} {et~al.}(1986){Kayser}, {Refsdal}, \&
  {Stabell}}]{1986AAKayser}
{Kayser}, R., {Refsdal}, S., \& {Stabell}, R. 1986, \aap, 166, 36

\bibitem[{{Kilic} {et~al.}(2007){Kilic}, {Allende Prieto}, {Brown}, \&
  {Koester}}]{2007ApJKilic}
{Kilic}, M., {Allende Prieto}, C., {Brown}, W.~R., \& {Koester}, D. 2007, \apj,
  660, 1451, \dodoi{10.1086/514327}

\bibitem[{{Kosakowski} {et~al.}(2022){Kosakowski}, {Kilic}, {Brown},
  {Bergeron}, \& {Kupfer}}]{2022MNRASKosakowski}
{Kosakowski}, A., {Kilic}, M., {Brown}, W.~R., {Bergeron}, P., \& {Kupfer}, T.
  2022, \mnras, 516, 720, \dodoi{10.1093/mnras/stac1146}

\bibitem[{{Kruse} \& {Agol}(2014)}]{KruseAgol2014}
{Kruse}, E., \& {Agol}, E. 2014, Science, 344, 275,
  \dodoi{10.1126/science.1251999}

\bibitem[{Liebes(1964)}]{Liebes1964}
Liebes, S. 1964, Phys. Rev., 133, B835, \dodoi{10.1103/PhysRev.133.B835}

\bibitem[{{Maeder}(1973)}]{1973AAMaeder}
{Maeder}, A. 1973, \aap, 26, 215

\bibitem[{{Masuda} {et~al.}(2019){Masuda}, {Kawahara}, {Latham}, {Bieryla},
  {Kunitomo}, {MacLeod}, \& {Aoki}}]{2019ApJLMasuda}
{Masuda}, K., {Kawahara}, H., {Latham}, D.~W., {et~al.} 2019, \apjl, 881, L3,
  \dodoi{10.3847/2041-8213/ab321b}

\bibitem[{{Narayan} \& {Bartelmann}(1996)}]{1996Narayan}
{Narayan}, R., \& {Bartelmann}, M. 1996, arXiv e-prints, astro,
  \dodoi{10.48550/arXiv.astro-ph/9606001}

\bibitem[{{Refsdal}(1964)}]{1964MNRASrefsdal}
{Refsdal}, S. 1964, \mnras, 128, 295, \dodoi{10.1093/mnras/128.4.295}

\bibitem[{Renn {et~al.}(1997)Renn, Sauer, \& Stachel}]{renn1997origin}
Renn, J., Sauer, T., \& Stachel, J. 1997, Science, 275, 184

\bibitem[{{Ricker} {et~al.}(2014){Ricker}, {Winn}, {Vanderspek}, {Latham},
  {Bakos}, {Bean}, {Berta-Thompson}, {Brown}, {Buchhave}, {Butler}, {Butler},
  {Chaplin}, {Charbonneau}, {Christensen-Dalsgaard}, {Clampin}, {Deming},
  {Doty}, {De Lee}, {Dressing}, {Dunham}, {Endl}, {Fressin}, {Ge}, {Henning},
  {Holman}, {Howard}, {Ida}, {Jenkins}, {Jernigan}, {Johnson}, {Kaltenegger},
  {Kawai}, {Kjeldsen}, {Laughlin}, {Levine}, {Lin}, {Lissauer}, {MacQueen},
  {Marcy}, {McCullough}, {Morton}, {Narita}, {Paegert}, {Palle}, {Pepe},
  {Pepper}, {Quirrenbach}, {Rinehart}, {Sasselov}, {Sato}, {Seager},
  {Sozzetti}, {Stassun}, {Sullivan}, {Szentgyorgyi}, {Torres}, {Udry}, \&
  {Villasenor}}]{Ricker2024}
{Ricker}, G.~R., {Winn}, J.~N., {Vanderspek}, R., {et~al.} 2014, in Society of
  Photo-Optical Instrumentation Engineers (SPIE) Conference Series, Vol. 9143,
  Space Telescopes and Instrumentation 2014: Optical, Infrared, and Millimeter
  Wave, ed. J.~{Oschmann}, Jacobus~M., M.~{Clampin}, G.~G. {Fazio}, \& H.~A.
  {MacEwen}, 914320, \dodoi{10.1117/12.2063489}

\bibitem[{{Riess} {et~al.}(1998){Riess}, {Filippenko}, {Challis},
  {Clocchiatti}, {Diercks}, {Garnavich}, {Gilliland}, {Hogan}, {Jha},
  {Kirshner}, {Leibundgut}, {Phillips}, {Reiss}, {Schmidt}, {Schommer},
  {Smith}, {Spyromilio}, {Stubbs}, {Suntzeff}, \& {Tonry}}]{1998AJRiess}
{Riess}, A.~G., {Filippenko}, A.~V., {Challis}, P., {et~al.} 1998, \aj, 116,
  1009, \dodoi{10.1086/300499}

\bibitem[{{Riess} {et~al.}(2024){Riess}, {Anand}, {Yuan}, {Casertano},
  {Dolphin}, {Macri}, {Breuval}, {Scolnic}, {Perrin}, \&
  {Anderson}}]{2024ApJRiess}
{Riess}, A.~G., {Anand}, G.~S., {Yuan}, W., {et~al.} 2024, \apjl, 962, L17,
  \dodoi{10.3847/2041-8213/ad1ddd}

\bibitem[{{Sahu} {et~al.}(2022){Sahu}, {Anderson}, {Casertano}, {Bond},
  {Udalski}, {Dominik}, {Calamida}, {Bellini}, {Brown}, {Rejkuba}, {Bajaj},
  {Kains}, {Ferguson}, {Fryer}, {Yock}, {Mr{\'o}z}, {Koz{\l}owski},
  {Pietrukowicz}, {Poleski}, {Skowron}, {Soszy{\'n}ski}, {Szyma{\'n}ski},
  {Ulaczyk}, {Wyrzykowski}, {Barry}, {Bennett}, {Bond}, {Hirao}, {Silva},
  {Kondo}, {Koshimoto}, {Ranc}, {Rattenbury}, {Sumi}, {Suzuki}, {Tristram},
  {Vandorou}, {Beaulieu}, {Marquette}, {Cole}, {Fouqu{\'e}}, {Hill}, {Dieters},
  {Coutures}, {Dominis-Prester}, {Bennett}, {Bachelet}, {Menzies}, {Albrow},
  {Pollard}, {Gould}, {Yee}, {Allen}, {Almeida}, {Christie}, {Drummond},
  {Gal-Yam}, {Gorbikov}, {Jablonski}, {Lee}, {Maoz}, {Manulis}, {McCormick},
  {Natusch}, {Pogge}, {Shvartzvald}, {J{\o}rgensen}, {Alsubai}, {Andersen},
  {Bozza}, {Novati}, {Burgdorf}, {Hinse}, {Hundertmark}, {Husser}, {Kerins},
  {Longa-Pe{\~n}a}, {Mancini}, {Penny}, {Rahvar}, {Ricci}, {Sajadian},
  {Skottfelt}, {Snodgrass}, {Southworth}, {Tregloan-Reed}, {Wambsganss},
  {Wertz}, {Tsapras}, {Street}, {Bramich}, {Horne}, {Steele}, \& {RoboNet
  Collaboration}}]{2022ApJSahu}
{Sahu}, K.~C., {Anderson}, J., {Casertano}, S., {et~al.} 2022, \apj, 933, 83,
  \dodoi{10.3847/1538-4357/ac739e}

\bibitem[{{Sajadian}(2021)}]{2021MNRAsajadian}
{Sajadian}, S. 2021, \mnras, 506, 3615, \dodoi{10.1093/mnras/stab1907}

\bibitem[{{Sajadian}(2023)}]{2023MNRASSajadian}
---. 2023, \mnras, 521, 6383, \dodoi{10.1093/mnras/stad945}

\bibitem[{sajadian(2024)}]{sajadian2024b}
sajadian, s. 2024, {Finite-Lens Effect on Self-Lensing in detached White
  Dwarfs-Main Sequence Binary Systems},  Zenodo,
  \dodoi{10.5281/zenodo.13851505}

\bibitem[{{Sajadian} \& {Afshordi}(2024)}]{Sajadian2024tess}
{Sajadian}, S., \& {Afshordi}, N. 2024, arXiv e-prints, arXiv:2409.12441,
  \dodoi{10.48550/arXiv.2409.12441}

\bibitem[{{Sajadian} \& {Hundertmark}(2017)}]{2017ApJsajadian}
{Sajadian}, S., \& {Hundertmark}, M. 2017, \apj, 838, 157,
  \dodoi{10.3847/1538-4357/aa67e1}

\bibitem[{{Sajadian} \& {Rahvar}(2010)}]{2010MNRAsajadian}
{Sajadian}, S., \& {Rahvar}, S. 2010, \mnras, 407, 373,
  \dodoi{10.1111/j.1365-2966.2010.16901.x}

\bibitem[{{Schneider} {et~al.}(1992){Schneider}, {Ehlers}, \&
  {Falco}}]{1992bookSchneider}
{Schneider}, P., {Ehlers}, J., \& {Falco}, E.~E. 1992, {Gravitational Lenses},
  \dodoi{10.1007/978-3-662-03758-4}

\bibitem[{{Schneider} \& {Weiss}(1987)}]{1987AASchneiderw}
{Schneider}, P., \& {Weiss}, A. 1987, \aap, 171, 49

\bibitem[{Soldner(1921)}]{soldner1921distraction}
Soldner, J. 1921, Annals of Physics, 370, 593

\bibitem[{{Sorabella} {et~al.}(2024){Sorabella}, {Laycock}, {Christodoulou}, \&
  {Bhattacharya}}]{2024ApJSorebella}
{Sorabella}, N.~M., {Laycock}, S. G.~T., {Christodoulou}, D.~M., \&
  {Bhattacharya}, S. 2024, \apjl, 961, L45, \dodoi{10.3847/2041-8213/ad19dc}

\bibitem[{{Vanderburg} {et~al.}(2020){Vanderburg}, {Rappaport}, {Xu},
  {Crossfield}, {Becker}, {Gary}, {Murgas}, {Blouin}, {Kaye}, {Palle}, {Melis},
  {Morris}, {Kreidberg}, {Gorjian}, {Morley}, {Mann}, {Parviainen}, {Pearce},
  {Newton}, {Carrillo}, {Zuckerman}, {Nelson}, {Zeimann}, {Brown},
  {Tronsgaard}, {Klein}, {Ricker}, {Vanderspek}, {Latham}, {Seager}, {Winn},
  {Jenkins}, {Adams}, {Benneke}, {Berardo}, {Buchhave}, {Caldwell},
  {Christiansen}, {Collins}, {Col{\'o}n}, {Daylan}, {Doty}, {Doyle},
  {Dragomir}, {Dressing}, {Dufour}, {Fukui}, {Glidden}, {Guerrero}, {Guo},
  {Heng}, {Henriksen}, {Huang}, {Kaltenegger}, {Kane}, {Lewis}, {Lissauer},
  {Morales}, {Narita}, {Pepper}, {Rose}, {Smith}, {Stassun}, \&
  {Yu}}]{2020NaturVanderburg}
{Vanderburg}, A., {Rappaport}, S.~A., {Xu}, S., {et~al.} 2020, \nat, 585, 363,
  \dodoi{10.1038/s41586-020-2713-y}

\bibitem[{{Wambsganss}(1998)}]{1998LRRWambsganss}
{Wambsganss}, J. 1998, Living Reviews in Relativity, 1, 12,
  \dodoi{10.12942/lrr-1998-12}

\bibitem[{{Witt} \& {Mao}(1994)}]{1994ApJWitt}
{Witt}, H.~J., \& {Mao}, S. 1994, \apj, 430, 505, \dodoi{10.1086/174426}

\bibitem[{{Yamaguchi} {et~al.}(2024){Yamaguchi}, {El-Badry}, {Ciardi},
  {Latham}, {Masuda}, {Bieryla}, {Clark}, \& {Condon}}]{2024Yamaguchi}
{Yamaguchi}, N., {El-Badry}, K., {Ciardi}, D.~R., {et~al.} 2024, \pasp, 136,
  074201, \dodoi{10.1088/1538-3873/ad5ebd}

\end{thebibliography}
\end{document}